\documentclass{JHEP3}

\usepackage{amsmath}
\usepackage{amsfonts}
\usepackage{amssymb}
\usepackage{amscd}
\usepackage{graphicx}
\usepackage{citesort}
\usepackage{mathrsfs}
\usepackage{bbm}
\usepackage{amsmath, amssymb, graphics}
\usepackage{amsmath,amssymb}
\usepackage{amsfonts,mathrsfs}
\usepackage{young}

\def\S{{\mathbb S}}

\newcommand{\diag}{{\rm diag}}
\newcommand{\fR}{{\mathbb R}}
\newcommand{\fL}{{\mathbb L}}
\newcommand{\fQ}{{\mathbb Q}}
\newcommand{\fC}{{\mathbb C}}
\newcommand{\fGG}{{\mathbb G}}
\newcommand{\fHH}{{\mathbb H}}

\newcommand{\gen}[1]{\mathfrak{#1}}

\def\[{\begin{equation}}
\def\]{\end{equation}}
\def\<{\begin{eqnarray}}
\def\>{\end{eqnarray}}
\def\beq{\begin{eqnarray}}
\def\eeq{\end{eqnarray}}

\newcommand{\half}{\frac{1}{2}}

\newcommand{\alg}[1]{\mathfrak{#1}}
\newcommand{\su}{\alg{su}}

\newcommand{\sls}{\alg{sl}}
\newcommand{\psl}{\alg{psl}}

\newcommand{\mathsym}[1]{{}}
\newcommand{\gl}{\alg{gl}}

\def\({\left(}
\def\){\right)}
\renewcommand{\[}{\left[}
\renewcommand{\]}{\right]}
\renewcommand{\eqref}[1]{$\({\rm \ref{#1}}\)$}

\numberwithin{equation}{section}

\def\[{\begin{equation}}
\def\]{\end{equation}}
\def\<{\begin{eqnarray}}
\def\>{\end{eqnarray}}

\usepackage{amsmath}
\usepackage{amsfonts}
\usepackage{amssymb}
\usepackage{amscd}
\usepackage{graphicx}
\usepackage{citesort}
\usepackage{mathrsfs}
\usepackage{bbm}

\def\S{{\mathbb S}}

\def\ads{{\rm AdS}_5\times {\rm S}^5}
\author{Gleb Arutyunov$^a$\footnote{ Correspondent fellow at
Steklov Mathematical Institute, Moscow.}, Marius de Leeuw$^a$\,  and\,  Alessandro Torrielli$^a$\footnote{Emails:
G.E.Arutyunov@uu.nl, M.deLeeuw@uu.nl,
A.Torrielli@uu.nl}
 \\
  \\ $^{a}$ {\it Institute for Theoretical
Physics and Spinoza Institute,\\ ~~Utrecht University, 3508 TD
Utrecht, The Netherlands} \\ }

\title{On Yangian and Long Representations of the Centrally Extended $\su(2|2)$ Superalgebra}

\abstract{The centrally extended $\su(2|2)$ superalgebra is an
asymptotic symmetry of the light-cone string sigma model on
$\ads$. We consider an evaluation representation of the
{\rm conventional} Yangian built over a particular 16-dimensional long
representation of the centrally extended $\su(2|2)$. Interestingly,
we find that {\rm S-matrices
compatible with this evaluation representation do not exist}. On the
other hand, {\rm by} requiring {\rm centrally extended}
$\su(2|2)$ invariance and explicitly {\rm solving} the Yang-Baxter
equation, we find {\rm a} scattering matrix
for long-short representations of the Lie superalgebra. {\rm We
notice that this S-matrix is invariant under a different representation of non-evaluation type, induced from the tensor
product of short representations. Our findings concern the conventional Yangian only, and are not applied to possible algebraic extensions of the latter.}}

\preprint{
          \tiny{ITP-UU--09-55}\\[-.5ex]
          \tiny{SPIN-09-45}\\[-.5ex]
          }

\begin{document}
\section{Introduction}
An important progress towards the solution of the finite-size
spectral problem of the planar AdS/CFT system has {\rm been} recently
made. On the one hand, a generalized L\"uscher approach for
treating the leading wrapping effects was developed
\cite{Ambjorn:2005wa,Bajnok:2008bm,Janik:2007wt,Hatsuda:2008gd,Minahan:2008re,Gromov:2008ie,Penedones:2008rv,Hatsuda:2008na,Bajnok:2008qj,Beccaria:2009eq,Beccaria:2009hg,Fiamberti:2009jw,Bajnok:2009vm}
and successfully confronted against direct field-theoretic
computations \cite{Fiamberti:2007rj,Velizhanin:2008jd}. On the
other hand, the Thermodynamic Bethe Ansatz based on the mirror
theory \cite{Arutyunov:2007tc} was advanced as a tool to capture
the exact string spectrum in both the 't Hooft coupling constant
$\lambda$ and the size $L$ of the system
\cite{Arutyunov:2009zu,Arutyunov:2009ur,Arutyunov:2009ux,Bombardelli:2009ns,Gromov:2009tv,Gromov:2009bc,Hegedus:2009ky,Frolov:2009in,Gromov:2009zb,Gromov:2009tq,Arutyunov:2009ax,Lukowski:2009ce,Arutyunov:2010gb,Balog:2010xa}.
In many respects the success of this research is based on the
existence of an asymptotic symmetry, which consists of (two
copies of) the $\su(2|2)$ superalgebra
\cite{Beisert:2005tm,Arutyunov:2006ak,Arutyunov:2009ga} centrally extended by two
central charges. In particular, this superalgebra and the
associated Yangian \cite{Beisert:2007ds} have been used to
explicitly determine the S-matrices describing scattering of
fundamental and bound-state particles of the light-cone string
sigma model \cite{Arutyunov:2009mi,Arutyunov:2009iq}, which is
important for setting up the Thermodynamic Bethe Ansatz approach.

\smallskip

Albeit nice, some of the recent developments were based on certain
assumptions and clever guesses, which provides us with further
motivation to better understand the representation theory of the
centrally extended $\su(2|2)$ superalgebra, as well as its
implications. So far mainly short (atypical) representations have
been the subject of interest in the context of the string sigma
model. This is because these representations correspond to bound
states of fundamental particles \cite{Dorey:2006dq} and, together
with the latter, they constitute the asymptotic spectrum of the
sigma model. On the other hand, long (typical) representations
naturally enter in the construction of the large $L$ asymptotic
solution of the TBA equations via the so-called Y-functions
\cite{Gromov:2009tv}. It is therefore interesting to look for {\rm at the}
scattering theory involving long representations. Another
independent aspect where long representations should play a  role
concerns the issue of the universal R-matrix. If such a quantity
exists as an abstract element in $\mathscr{A}\otimes \mathscr{A}$,
where $\mathscr{A}$ is a Hopf algebra, then it can be evaluated in
any two representation of $\mathscr{A}$. In the case of (the
Yangian of) centrally-extended $\su(2|2)$, these must include the
long ones. Therefore, there should exist a concrete (matrix)
realization for an intertwiner of symmetry generators in the
tensor product of the corresponding (long) representations (the
S-matrix).

\smallskip

In this paper, we will make a first step towards understanding the
scattering problem involving long representations of the centrally
extended $\su(2|2)$. We start with building such long
representations by applying an outer $\alg{sl}(2)$ automorphism to
the representations of the unextended $\su(2|2)$ superalgebra
\cite{Beisert:2006qh}. These representations, in turn, can be
obtained {\rm from} those constructed for $\alg{gl}(2|2)$ by Gould
and Zhang \cite{zhang-2005-46}, {\rm see also
\cite{Kamupingene:1989wj,Palev:1990wm}}. They are parameterized by
a continuous parameter $q\in \mathbb{C}$, which is the value of
the unique central charge (the Hamiltonian) in a given
representation. An outer $\alg{sl}(2)$ automorphism acting on
$\su(2|2)$ can be used to generate two extra central charges,
which depend on additional parameters $P$ and $g$. Here $P$ is
identified with the (generically complex) particle momentum, while
$g$ with the coupling constant. We will be interested in the
lowest (16-dimensional) long representation, for which we will
obtain an explicit realization in terms of $16\times 16$ matrices
depending on $q,P$ and $g$. As usual, special values of $q$
correspond to the shortening conditions. In particular, $q=1$
corresponds to an indecomposable formed out of two short
8-dimensional representations.

\smallskip

Given an explicit realization of the long 16-dimensional
representation, we construct the corresponding evaluation
representation for the Yangian introduced by Beisert
\cite{Beisert:2007ds}. {\rm The defining relations are given in Appendices \ref{App;Yangian} and \ref{App:DrinII}. We will refer to this Yangian, exclusively built upon the centrally extended $\su(2|2)$ superalgebra, as the {\it conventional} Yangian. Whenever the term `Yangian' will be used throughout the paper, it will always be understood as conventional, even if we will not mention it explicitly.

We will then} try to find an S-matrix which
scatters the long {\rm evaluation} with the long {\rm evaluation} or the long {\rm evaluation} with a short
four-dimensional representation. {\rm Namely}, we try to find an S-matrix which acts as the
following intertwiner: \beq\nonumber
\Delta^{op}(\mathbb{J})\,\mathbb{S}=\mathbb{S}\,\Delta(\mathbb{J})\,
, ~~~~~~~
\Delta^{op}(\, \widehat{\mathbb{J}}\, )\,\mathbb{S}=\mathbb{S}\,\Delta(\, \widehat{\mathbb{J}} \,)\,
, \eeq where $\mathbb{J}$ is a generator of {\rm centrally extended} $\su(2|2)$,
$\widehat{\mathbb{J}}$ is the corresponding Yangian generator in the
evaluation representation, and $\Delta$ and $\Delta^{op}$ are the
coproduct and its opposite (see section 2 for the precise
definitions). We recall that this construction proved to work well
for the fundamental or bound-state, {\it i.e.} short,
representations, and it {\rm lead} to the complete determination of the
corresponding bound state scattering matrix \cite{deLeeuw:2008dp,Arutyunov:2009mi}. However,
where one of the representations involved is long {\rm evaluation}, we find that
the S-matrix satisfying the invariance conditions above {\it does not
exist}.

The origin of this problem can be {\rm clearly seen in Drinfeld's second realization \cite{Spill:2008tp}\footnote{\rm Given the existence of an invertible map between the generators of Drinfeld's second \cite{Spill:2008tp} and first \cite{Beisert:2007ds} realization of the ${}$ Yangian, we will use either realizations according to the needs considering them as completely equivalent. A rigorous proof of this equivalence has however never been derived.}, where it can be} traced back to
non-co-commutativity of the higher Yangian central charges
$\mathbb{C}_n$ with $n\geq 2$. {\rm {\rm Co-commutativity} means that the coproduct map coincides with its opposite, namely $\Delta = \Delta^{op}$. When a coproduct will be called co-commutative without further specification, it means it is such when it acts on all the generators of the algebra in question. We will often use the terminology {\rm co-commutativity of the generator $X$} as a shortcut for {\rm co-commutativity of the coproduct when acting on generator $X$}, namely $\Delta (X) = \Delta^{op} (X)$. The term {\rm quasi-co-commutative} referred to a coproduct means that the coproduct map is equal to its opposite up to conjugation with an {\rm invertible} element $\mathbb{S}$, namely $ \Delta^ {op} \, \mathbb{S} = \mathbb{S} \, \Delta$. This fact can happen in some representations, and not in others. If it happens in all representations, and the element $\mathbb{S}$ can be determined in a representation-independent way as an abstract object, we will call it the {\rm universal R-matrix}. If {\rm only} some representations allow the coproduct to be related to its opposite by conjugation with a invertible matrix $\mathbb{S}$, we will say that those representations {\rm admit an S-matrix}, but there is no universal R-matrix.

Since the coproducts of the Yangian central charges {\rm only} involve central elements (see formula (\ref{eqn;CoProdYangDiff})), co-commutativity of the central charges in a specific representation is a
necessary condition for the existence of an S-matrix in that representation. In fact, one has
\begin{eqnarray}
\Delta^{op}(\mathbb{C}_n)\,\mathbb{S}=\mathbb{S}\,\Delta^{op}(\mathbb{C}_n)=\mathbb{S}\,\Delta(\mathbb{C}_n)
\end{eqnarray}
from which invertibility of $\mathbb{S}$ implies $\Delta^{op}(\mathbb{C}_n)=\Delta(\mathbb{C}_n)$.
Finding at least one representation of the Yangian where this necessary condition is not satisfied\footnote{For instance, if one takes the coproduct of $\mathbb{C}_2$ in a {\it long} $\otimes$ {\it long} representation, and subtracts from it its opposite, one obtains an expression which expands  semiclassically as $$\Delta (\mathbb{C}_2 ) - \Delta^{op} (\mathbb{C}_2 ) = \frac{x_1 x_2 q_1 q_2 (q_2^2 - q_1^2)}{2 g (x_1^2 - 1)(x_2^2 - 1)} + {\cal{O}} (g^{-2}),$$ with $x$ a classical rapidity.} implies that the corresponding universal
R-matrix does not exist.

\smallskip

Even if the Yangian evaluation representation does not admit an S-matrix}, one can still look for
{\rm centrally extended} $\su(2|2)$-invariant solutions of the Yang-Baxter equation.
Indeed, we find two such solutions. It is important to understand if there is still some
extended symmetry they correspond to. To answer this question,
we recall that, generically, a product of two short representations
gives an {\it irreducible} long representation \cite{Beisert:2006qh}. In
particular,
$$
V_{4d}(p_1)\otimes V_{4d}(p_2) \approx V_{16d}(P,q)\, .
$$
Here, $V_{4d}(p)$ is a  fundamental 4-dimensional representation
which depends on the particle momentum and the coupling constant.
Analogously, $V_{16d}(P,q)$ is a long 16-dimensional
representation described by the momentum $P$, the coupling
constant $g$  and the parameter $q$. We find an explicit relation
between the pairs $(p_1,p_2)$ and $(P,q)$ at fixed $g$, in
particular $P=p_1+p_2$. Obviously, for a given $p_1$ and $p_2$
there is a unique long representation. However, any long
representation can be written as a tensor product of two short
representations in two different ways.

The observed relationship between long and short representations
suggests that the S-matrix $\mathbb{S}_{LS}$, which scatters a
long representation with a short one, can simply be composed as a
product of two S-matrices $\mathbb{S}_{13}$ and $\mathbb{S}_{23}$ describing the
scattering of the corresponding short representations, {\it i.e.}
$$
{\mathbb S}_{LS}(P,q;p_3)={\mathbb S}_{13}(p_1,p_3){\mathbb
S}_{23}(p_2,p_3)\, .
$$
In this formula, the tensor product of two short representations
in the spaces $1$ and $2$ with momenta $p_1$ and $p_2$ gives a
long representation $(P,q)$, which scatters with a short
representation in the third space with momentum $p_3$. We then
verify that the two S-matrices we found by solving the Yang-Baxter
equations indeed coincide with the product of two ``short"
S-matrices. The fact that we find two matrices is explained by the
double-covering relationship between $(p_1,p_2)$ and $(P,q)$. This
finding also shows that the {\rm ${}$} Yangian symmetry can be induced on
long representation from the one defined on the short ones, {\rm and this tensor product representation}
automatically {\rm admits an S-matrix. It naturally does that in both branches of the double-covering. Importantly, this (double-branched) tensor product representation of the ${}$ Yangian is {\it not} isomorphic to the long evaluation representation we were discussing before, even though the two short representations composing it are the short evaluation representations of the ${}$ Yangian described in \cite{Beisert:2007ds,Spill:2008tp}. This is also clear from the fact that the long evaluation representation discussed before {\it does not} admit an S-matrix.}

The existence of {\rm {\it two} solutions for ${\mathbb S}_{LS}$,
corresponding to two Yangian representations} induced from short
representations, is an unexpected feature which we do not have a
good explanation for. Both S-matrices come with the canonical
normalization and, therefore, they cannot be related to each other
by any multiplicative factor (an extra dressing phase). They are
not related by a similarity transformation either. We note,
however, that at the special value $q=1$ where the long multiplet
becomes reducible, the two matrices ${\mathbb S}_{LS}$ become of
the form (the block structure refers to the split into the
8-dimensional sub- and factor representations one finds at $q=1$,
as we will discuss in the paper)
\begin{align}
\begin{pmatrix}
 \alpha A & \, \, B + \alpha C \\
 0 & D
\end{pmatrix},
\end{align}
where only the scalar coefficient $\alpha$ is different for the two solutions. Here, $D$ corresponds to the factor representation (symmetric), and coincides with (the inverse of) the known symmetric bound-state S-matrix $\S_{AB}$ \cite{Arutyunov:2008zt}. This is in
agreement with the fact that there is a unique bound-state
S-matrix.

\smallskip

The paper is organized as follows. In the next section, starting
from a construction of long representations, we discuss the {\rm
${}$} Yangian and prove the {\rm non-existence of a universal
R-matrix. This no-go theorem applies to the conventional Yangian
only, and it may not hold when considering algebraic extensions of
the latter\footnote{\rm In fact, certain extensions may result in
further relations one has to impose on the generators. These extra
relations may not be satisfied by the evaluation representation,
therefore ruling it out from the list of irreps. We thank Niklas
Beisert for a discussion about this point.}.} In section 3 we find
the ``long-short" S-matrix by solving the corresponding
Yang-Baxter equation. In section 4 we present an alternative
construction of long representations and the associated S-matrix
{\it via} tensor product of short ones. In appendices 5.1-5.3 we
provide several computational details. Finally, in appendix 5.4 we
discuss some aspects of the Hirota equations related to the long
representations we construct in the paper. Most of the
corresponding discussion should be known to experts, and we
include it only for completeness.

\section{Long representations and Yangian}
We start with discussing the representation theory of
$\alg{sl}(2|2)$ and its generalization to the centrally extended
case. We provide a matrix realization of the simplest
16-dimensional long multiplet. Then we discuss the evaluation
representation of the corresponding Yangian algebra based on this
long multiplet and show the absence of {\rm a universal R-matrix}.

\subsection{Constructing long representations}
The paper \cite{zhang-2005-46} explicitly constructs all
finite-dimensional irreducible representations of $\gl (2|2)$ in
an oscillator basis. Generators of $\gl (2|2)$ are denoted by
$E_{ij}$, with commutation relations \beq
[E_{ij},E_{kl}]=\delta_{jk} E_{il} - (-)^{(d[i]+d[j])(d[k]+d[l])}
\delta_{il} E_{kj}. \eeq Indices $i,j,k,l$ run from $1$ to $4$,
and the fermionic grading is assigned as $d[1]=d[2]=0$,
$d[3]=d[4]=1$. The quadratic Casimir of this algebra is $C_2 =
\sum_{i,j=1}^4 (-)^{d[j]} E_{ij}E_{ji}$. Finite dimensional irreps
are labelled by two half-integers $j_1,j_2 = 0,\frac{1}{2},...$,
and two complex numbers $q$ and $y$. These numbers correspond to
the values taken by appropriate generators on the highest weight
state $|\omega\rangle$ of the representation, defined by the
following conditions: \beq \label{zghw} &&H_1 |\omega\rangle = (E_{11}-E_{22})
|\omega\rangle = 2 j_1 |\omega\rangle, \qquad
H_2 |\omega\rangle = (E_{33}-E_{44}) |\omega\rangle = 2 j_2 |\omega\rangle,\nonumber\\
&&I |\omega\rangle = \sum_{i=1}^4 E_{ii} |\omega\rangle = 2 q
|\omega\rangle, \, \, \, N |\omega\rangle = \sum_{i=1}^4 (-)^{[i]}
E_{ii} |\omega\rangle = 2 y |\omega\rangle, \, \, \, \,
E_{i<j}|\omega\rangle =0. \eeq The generator $N$ never appears on
the right hand side of the commutation relations, therefore it is
defined up to the addition of a central element $\beta I$, with
$\beta$ a constant\footnote{We decided to drop the term $\beta I$
since it will not affect our discussion.}. This also means that we
can consistently mod out the generator $N$, and obtain $\sls
(2|2)$ as a subalgebra of the original $\gl (2|2)$
algebra\footnote{Further modding out of the center $I$ produces
the simple Lie superalgebra $\psl(2|2)$. Its representations can
be understood as that of $\alg{sl}(2|2)$ for which $q=0$.
Correspondingly, $\sl(2|2)$ has long irreps of dimension
$16(2j_1+1)(2j_2+1)$ with $j_1\neq j_2$ and short irreps with
$j_1=j=j_2$ of dimension $16j(j+1)+2$. For a discussion of the
tensor product decomposition of $\psl(2|2)$, see
\cite{Gotz:2005ka}.}. In order to construct representations of the
centrally-extended $\su (2|2)$ Lie superalgebra\footnote{The
reality condition on the algebra will be imposed later, and will
not affect the present discussion.}, we then first mod out $N$,
and subsequently perform an $\sls (2)$ rotation by means of the
outer automorphism of $\su (2|2)$ \cite{Beisert:2006qh}.

As usual for superalgebras, irreps are divided into typical
(long), which have generic values of the labels $j_1,j_2,q$, and
atypical (short), for which special relations are satisfied by
these labels. Short representations occur here for $\pm q = j_1 -
j_2$ and $\pm q = j_1 + j_2 +1$. When these relations are
satisfied, the dimension of the representation is smaller than
what it would generically be for the same values of $j_1,j_2$, but
$q$ arbitrary (that is, the dimension is in these cases smaller
than $16(2 j_1 +1)(2 j_2 +1)$). One notices also that, when
starting from a long irrep and reaching these special values by
continuous variation of the parameter $q$, one generically ends up
into a reducible but indecomposable representation.

We can identify the values of the labels which will produce the
representations we are particularly interested in in this paper.
First of all, the fundamental $4$-dimensional short representation
\cite{Beisert:2005tm} corresponds to $j_1=\frac{1}{2},j_2=0$ (or,
equivalently, $j_1=0,j_2=\frac{1}{2}$) and $q=\frac{1}{2}$
($q=-\frac{1}{2}$). More generally, the bound state (symmetric
short) representations
\cite{Dorey:2006dq,Chen:2006gp,Chen:2006gq,Roiban:2006gs,Beisert:2006qh,Arutyunov:2008zt}
are given by $j_2=0,q=j_1$, with $j_1 =\frac{1}{2},1,...$ and
bound state number $M \equiv s = 2 j_1$. In addition, there are
the antisymmetric short representations given by $j_1=0,q=1+j_2$,
with $j_2 =0,\frac{1}{2},...$ and bound state number $M\equiv
a =2(j_2 + 1)$. Both symmetric and antisymmetric representations have
dimension $4M$. We see that symmetric and antisymmetric
representations are associated with the different shortening
conditions $\pm q=j_1-j_2$ and $\pm q=1+j_1+j_2$.

Second, we consider the simplest long representation of dimension
16. In terms of the $\gl (2|2)$ labels introduced above, this is
the $16$-dimensional long representation characterized by
$j_1=j_2=0$, and arbitrary $q$. It is instructive to see how it
branches under the $\su (2) \oplus \su (2)$ algebra. We denote as
$[l_1,l_2]$ the subset of states which furnish a representation of
$\su (2) \oplus \su (2)$ with angular momentum $l_1$ w.r.t the
first $\su (2)$, and $l_2$ w.r.t the second $\su (2)$,
respectively. The branching rule is \beq \label{split} (2,2) \,
\rightarrow \, 2 \times [0,0] \oplus 2 \times
[\frac{1}{2},\frac{1}{2}] \oplus [1,0] \oplus [0,1]. \eeq One can
straightforwardly verify that the total dimension adds up to $16$,
since $[l_1,l_2]$ has dimension $(2 l_1 + 1)\times (2 l_2 + 1)$.

For generic values of $q$, the corresponding long representations
have no interpretation in terms of Young tableaux. However, when
$q$ is a certain integer, such an interpretation becomes possible.
Consider rectangular Young tableaux, with one side made of $2$
boxes, and the other side made of arbitrarily many boxes. These
are long representations, denoted by $(2,s)$ and $(a,2)$ according
to the length (in boxes) of their sides. Together with the short
irreps, denoted accordingly as $(1,s)$ (symmetric) and $(a,1)$
(antisymmetric), they span all the admissible rectangular
representations. In fact, every allowed representation has to have
its associated Young tableaux fit into the so-called ``fat hook"
\cite{Kazakov:2007na}, which has branches of width equal to two
boxes. All representations $(2,s)$ (respectively, $(a,2)$) with
$s\geq 2$ (respectively, $a\geq 2$) have
dimension\footnote{Formulas for computing the dimension of
representations of superalgebras from their Young tableaux can be
found in \cite{BaBa}.} 16, central charge $q=s$ and Dynkin labels
$[0,q,0]$. For both long and short representations that
have an interpretation in terms of a rectangular Young tableaux,
the charge $q$ is simply given by the number of boxes in the
tableaux multiplied by the charge $q=1/2$ of the fundamental
representation.

As a first step of our study, we have explicitly constructed the
oscillator representation by using the formulas of
\cite{zhang-2005-46}, and derived from it the $16 \times 16$
matrix realization of the algebra generators. We have done this
before acting with the outer automorphism, in such a way that the
subsequent $\sls(2)$ rotation provides an explicit matrix
representation of centrally-extended $\su (2|2)$. This explicit
realization is reported in appendix \ref{app:param}. Below we
discuss some of the salient features of this realization.

The way the outer automorphism is implemented is by mapping the
$\gl (2|2)$ non-diagonal generators into new generators as
follows: \beq &&\fL^b_a = E_{a b} \, \, \, \forall \, \, a \neq b,
\qquad \fR^\beta_\alpha = E_{\alpha \beta} \, \, \,
\forall \, \, \alpha \neq \beta,\nonumber\\
&&\fQ^a_\alpha = a \, E_{\alpha a} + b \, \epsilon_{\alpha \beta} \epsilon^{a b} E_{b \beta},\nonumber\\
&&\fGG^\alpha_a = c \, \epsilon_{a b} \epsilon^{\alpha \beta}
E_{\beta b} + d \, E_{a \alpha}, \eeq subject to the constraint
\beq ad - bc = 1. \eeq Diagonal generators are automatically
obtained by commuting positive and negative roots. In particular,
from the explicit matrix realization one obtains the following
values of the central charges: \beq \fHH = 2 q \, (a d + b c) \,
\mathbbmss{1}, \qquad \fC = 2 q \, a b \, \mathbbmss{1}, \qquad
\fC^\dagger = 2 q \, c d \, \mathbbmss{1}, \eeq ($\mathbbmss{1}$
is the $16$-dimensional identity matrix), satisfying the condition
\beq \label{conditio} \frac{\fHH^2}{4} - \fC \fC^\dagger = q^2 \,
\mathbbmss{1}. \eeq When $q^2 =1$, this becomes a shortening
condition. In fact, for $q =1$ the $16$-dimensional representation
becomes reducible but indecomposable. Its subrepresentation
\cite{Gotz:2005ka} is a short anti-symmetric $8$-dimensional
representation. Formula (\ref{conditio}) above, however, tells us
that we can conveniently think of $q$ as a {\it generalized} bound
state number, since for short representations $2 q$ would be
replaced by the bound state number $M$ in the analogous formula
for the central charges. This is particularly useful, since it
allows us to parameterize the labels $a,b,c,d$ in terms of the
familiar bound state variables\footnote{We use the conventions of
\cite{Arutyunov:2009mi}.} $x^\pm$, just replacing the bound state
number $M$ by $2 q$. The explicit parameterization is given by
\begin{align}
\label{param}
a &= \sqrt{\frac{g}{4q}}\eta,  &b &=-\sqrt{\frac{g}{4q}}\frac{i}{\eta}\left(1-\frac{x^+}{x^-}\right) ,\nonumber\\
c &=-\sqrt{\frac{g}{4q}} \frac{\eta}{x^+}  , &d&= \sqrt{\frac{g}{4q}}\frac{x^+}{i\eta}\left(1-\frac{x^-}{x^+}\right),
\end{align}
where
\begin{align}
\label{eta} \eta = e^{\frac{ip}{4}}\sqrt{i(x^- - x^+)}
\end{align}
and
\begin{align}
x^+ + \frac{1}{x^+} - x^- - \frac{1}{x^-} \, = \, \frac{4 i q}{g}.
\end{align}
As in the case of short representations, there exist a
uniformizing torus with variable $z$ and periods depending on $q$
\cite{Janik:2006dc} . The choice (\ref{eta}) for $\eta$ is
historically preferred in the string theory analysis
\cite{Arutyunov:2006yd,Arutyunov:2007tc,Arutyunov:2008zt,Arutyunov:2009mi},
and will actually ensure our S-matrix to be symmetric. Finally, we
point out that positive and negative values of $q$ correspond to
positive and negative energy representations, respectively.

\subsection{Hopf algebra and Yangian}\label{sect:hy}

Having in mind the derivation of an S-matrix in the above
described long representation, we equip the symmetry algebra with
the deformed Hopf-algebra coproduct\footnotemark[\value{footnote}]
\cite{Gomez:2006va,Plefka:2006ze}
\beq
\label{cop}
&&\Delta (\mathbb{J}) = \mathbb{J} \otimes \mathbb{U}^{[[\mathbb{J}]]} +
\mathbbmss{1} \otimes \mathbb{J}, \nonumber\\
&&\Delta(\mathbb{U})=\mathbb{U}\otimes \mathbb{U}, \eeq where
$\mathbb{J}$ is any generator of centrally-extended $\su (2|2)$,
$[[\mathbb{J}]]=0$ for the bosonic $\su (2) \oplus \su (2)$
generators and for the energy generator $\mathbb{H}$,
$[[\mathbb{J}]]=1$ (resp., $-1$) for the $\mathbb{Q}$ (resp.,
$\mathbb{G}$) supercharges, and $[[\mathbb{J}]]=2$ (resp., $-2$)
for the central charge $\mathbb{C}$ (resp., $\mathbb{C}^\dagger$).
The value of $\mathbb{U}$ is determined by the consistency
requirement that the coproduct is co-commutative on the center,
which is a necessary condition for the existence of an S-matrix
$\S$ satisfying\footnote{$\S$ acts as $\S : V_1 \otimes V_2
\rightarrow V_1 \otimes V_2$ on representation modules.} \beq
\label{coprod} \Delta^{op} (\mathbb{J}) \, \, \S = \S \, \Delta
(\mathbb{J}) \qquad \forall \, \mathbb{J}. \eeq This produces the
algebraic condition \beq \mathbb{U}^2 \, = \, \kappa \, \mathbb{C}
\, + \, \mathbbmss{1} \eeq for some representation-independent
constant $\kappa$. With our choice of parametrization
(\ref{param}), $\kappa$ gets re-expressed {\it via} the coupling
constant $g$ as $\kappa = \frac{2}{i g}$, and we obtain the
familiar relation \beq \mathbb{U} \, = \, \sqrt{\frac{x^+}{x^-}}
\, \mathbbmss{1} \, =\, e^{i \frac{p}{2}} \, \mathbbmss{1}. \eeq
The advantage of the choice (\ref{param}) is that the above
relations are valid as they stand both for long and short
representations. This will be particularly useful, since we plan
to project the coproduct (\ref{cop}) into a long representation in
the first space, and a short one in the second space. This is so
because we will be primarily interested in the S-matrix scattering
long representations against short ones.

In order to have a complete realization of the Hopf algebra, one
needs to remember the antipode map $\cal{S}$
\cite{Janik:2006dc,Plefka:2006ze,Arutyunov:2008zt}, and specify
the charge conjugation matrix $\cal{C}$ that implements the
antipode in the long representations\footnote{The counit and all
other bialgebra structures are straightforwardly implemented, and do
not present any novel features.}. One has in particular \beq
\label{anti} {\cal{S}} (\mathbb{J}) \, = \, -
\mathbb{U}^{-[[\mathbb{J}]]} \, \mathbb{J} \, = \, {\cal{C}} \,
\overline{\mathbb{J}}^{st} \, {\cal{C}}^{-1}, \eeq where
$\overline{\mathbb{J}}$ is the {\it antiparticle} representation
associated to the representation we choose on the l.h.s. of
(\ref{anti}). One finds that the charge conjugation matrix in the
$16$-dimensional long representation is given by

\begin{align}
\label{caric}
\mathcal{C}= \begin{pmatrix}
\fbox{ $\begin{smallmatrix} -1 \end{smallmatrix}$} & ~ & ~ & ~ & ~ & ~ \\
~ & \fbox{ $\begin{smallmatrix}
 ~ & ~ & ~ & i  \\
 ~ & ~ & -i & ~  \\
 ~ & -i & ~ & ~ \\
 i & ~ & ~ & ~
 \end{smallmatrix}$} & ~ & ~ & ~ & ~ \\
 ~ & ~ & \fbox{ $\begin{smallmatrix}
 ~ & ~ & -1  \\
 ~ & 1 & ~  \\
 -1 & ~ & ~
 \end{smallmatrix}$} & ~ & ~ & ~  \\
  ~ & ~ & ~ & \fbox{ $\begin{smallmatrix}
 ~ & ~ & -1  \\
 ~ & 1 & ~  \\
 -1 & ~ & ~
 \end{smallmatrix}$} & ~ & ~  \\
 ~ & ~ & ~ & ~ & \fbox{ $\begin{smallmatrix}
 ~ & ~ & ~ & i  \\
 ~ & ~ & -i & ~  \\
 ~ & -i & ~ & ~ \\
 i & ~ & ~ & ~
 \end{smallmatrix}$} & ~ \\
 ~ & ~ & ~ & ~ & ~ & \fbox{ $\begin{smallmatrix} -1 \end{smallmatrix}$}
\end{pmatrix}
\end{align}

\smallskip

The blocks in (\ref{caric}) refer to the branching rule (\ref{split}), with the ordering of states given in section \ref{kinstr}.
The antiparticle representation $\overline{\mathbb{J}}$ is still
defined by sending $p\rightarrow -p$ (together with changing sign to the eigenvalue of the energy generator $\mathbb{H}$), which means
\begin{eqnarray}
x^{\pm}\rightarrow \frac{1}{x^{\pm}},
\end{eqnarray}
exactly as in the case of short representations. Once again, on
the uniformizing torus (cf. comment to (\ref{param})), applying
the particle to anti-particle transformation four times gives the
identity, which corresponds to the $\mathbb{Z}_4$ graded
Lie-algebra structure of centrally extended $\alg{su}(2|2)$.

The next step is to study the Yangian in this representation. One
can prove that the defining commutation relations of Drinfeld's first
realization of the Yangian given in \cite{Beisert:2007ds} (see appendix \ref{App;Yangian}) are
satisfied (by the generators and their coproducts) if we assume the evaluation representation\footnote{We
use the conventions of \cite{Arutyunov:2009mi}.} \beq \label{eval}
\widehat{\mathbb{J}} \, = \, u \, \mathbb{J}, \eeq where the
spectral parameter $u$ assumes the familiar form \beq \label{u}
u \, = \,
\frac{g}{4 i} (x^+ + x^-)\bigg( 1 + \frac{1}{x^+ x^-} \bigg). \eeq The
Yangian coproducts are given by the same formulas used in
\cite{Arutyunov:2009mi}, and the above value of $u$ is determined
by requiring co-commutativity of the Yangian central charges
$\widehat{\mathbb{C}}$, $\widehat{\mathbb{C}}^\dagger$. We report the details in appendix \ref{App;Yangian} for convenience of the reader.

Drinfeld's second realization is also obtained by applying a similar (Drinfeld's) map as in \cite{Spill:2008tp}\footnote{We have checked that the map we use in this paper (see appendix \ref{App:DrinII}) also works for the fundamental representation equally well, and is, in this sense, universal. This map might be related to the one used in \cite{Spill:2008tp} by redefinitions of the generators in the various realizations.}. This ensures the
fulfilment of the Serre relations (see also \cite{Matsumoto:2009rf}). All defining relations in
\cite{Spill:2008tp} are satisfied (see appendix \ref{App:DrinII}), although the representation one
obtains after Drinfeld map is not any longer of a simple
evaluation-type, but more complicated. In fact, level-$n$ simple
roots $\mathbb{J}_n$ are not obtained from level-zero ones {\it
via} multiplication by a (possibly shifted) spectral parameter to
the power $n$. Nevertheless, the representation we obtain for Drinfeld's second
realization of the Yangian is consistent, and the coproducts obtained after
Drinfeld's map respect all commutation and Serre
relations\footnote{Antipode and charge conjugation are also
perfectly consistent with Drinfeld's second realization.}. We give details of this realization in appendix \ref{App:DrinII}.

However, surprisingly, it turns out that the Yangian in this
representation, both for coproducts
projected into {\it long} $\otimes$ {\it short} and for {\it long} $\otimes$ {\it long}
representations, {\rm does {\it not} admit an S-matrix}. This is easily seen by considering the
Yangian central charges $\mathbb{C}_n$, $\mathbb{C}^\dagger_n$. While for
$n=0,1$, their coproducts are co-commutative, this is not
so for $n\geq 2$. Only for the special case $q^2 =1$ the Yangian
central charges appear to be co-commutative also for $n=2$ and
higher\footnote{For these special values of $q$ we actually
checked co-commutativity only up to $n=4$.}. Nevertheless, even
for the special case $q^2 =1$, the Yangian still does not seem to
admit an S-matrix in this representation. One way to see it is by noticing that the
equation \beq \label{coprodo} \Delta^{op} ( \, \widehat{\mathbb{J}}
\, ) \, \, \S = \S \, \Delta ( \, \widehat{\mathbb{J}} \, ), \eeq
when applied to certain combinations of generators and on
particular states (for instance, of highest weight w.r.t. to the $\su (2) \oplus
\su (2)$ splitting (\ref{split})), leads to a contradiction when
the explicit matrix realization is used. This means that such an
S-matrix does not exist for this representation of the Yangian, which also implies
that a universal R-matrix for the
Yangian \cite{Beisert:2007ds} does not exist.

However, as we will discuss in section \ref{sect:short}, a
different Yangian representation,  for which an S-matrix does
indeed exists, can be induced on the space of long
representations. {\rm This Yangian representation is obtained {\it via} the decomposition of long representations into
short ones, and is therefore built upon the Yangian representations that have been
already built on short representations}. This induced
representation is quite different from the one described above
(cf. (\ref{eval})), and, in particular, it is not related to
(\ref{eval}) {\it via} any similarity transformation combined with
redefinition of the spectral parameters.

A remark is in order. In principle, it should be possible to deduce non-co-commutativity of the higher central charges directly from the corresponding formulas for the coproducts written in terms of algebra generators, without referring to a specific representation. These formulas should also imply that the non-co-commutative part must disappear for representations which satisfy the shortening conditions. However, the abstract formulation of the coproducts is quite cumbersome, and we find it more illuminating to exhibit a concrete representation for which the higher central charges show non-co-commutativity, cf.
footnote 2.

\section{The Long-Short S-matrix}

{\rm Let us start by deriving all S-matrices}
satisfying the relation (\ref{coprod}) exclusively for the
(level-zero) algebra generators.
{\rm Such an S-matrix describes} the scattering of an
excitation in the long representation with momentum $P$ and
parameter $q$, against a fundamental particle with momentum $p$:
\begin{align}
\S: V_{16d}(P,q)\otimes V_{4d}(p) \longrightarrow
V_{16d}(P,q)\otimes V_{4d}(p).
\end{align}
This S-matrix should relate the Hopf algebra structure to the
opposite Hopf algebra one. This means that $\S$ should satisfy (\ref{coprod}).

\subsection{Kinematic Structure}\label{kinstr}

It is useful to apply a procedure similar to the one performed in
\cite{Arutyunov:2009mi}, this time without the explicit help of
Yangian symmetry. We again begin by using the $\alg{su}(2) \oplus
\su (2)$ invariance (with trivial coproduct) to divide the space
$V_{16d}(P,q)\otimes V_{4d}(p)$ into blocks with definite
$\alg{su}(2)_{\mathbb{L}} \oplus \su (2)_{\mathbb{R}}$ weights
$(a;b)$. Let us denote the basis-vectors of the long
representation by $f_i$ and the basis for the short representation
by $e_i$, respectively. They correspond to $16$-dimensional
(resp., $4$-dimensional) vectors with all zeroes, except in
position $i$, where there is a $1$. The $\alg{su}(2)_{\mathbb{L}}
\oplus \su (2)_{\mathbb{R}}$ weights are explicitly given in
the following tables:

\bigskip

\hspace{-.7cm}\begin{tabular}{|c||c|c|c|c|c|c|c|c|c|c|c|c|c|c|c|c|}
  \hline $~$ & $f_1$ & $f_2$ & $f_3$ & $f_4$ & $f_5$ & $f_6$ & $f_7$ & $f_8$ & $f_9$ & $f_{10}$ & $f_{11}$ & $f_{12}$ & $f_{13}$ & $f_{14}$ & $f_{15}$ & $f_{16}$ \\
  \hline
  $\mathbb{L}^1_1$ & $0$ & $\frac{1}{2}$ & $-\frac{1}{2}$ & $\frac{1}{2}$ & $-\frac{1}{2}$ & $1$ & $0$ & $-1$ & $0$ & $0$ & $0$ & $\frac{1}{2}$ & $-\frac{1}{2}$ & $\frac{1}{2}$ & $-\frac{1}{2}$ & $0$ \\
  $\mathbb{R}^3_3$ & $0$ & $\frac{1}{2}$ & $\frac{1}{2}$ & $-\frac{1}{2}$ & $-\frac{1}{2}$ & $0$ & $0$ & $0$ & $1$ & $0$ & $-1$ & $\frac{1}{2}$ & $\frac{1}{2}$ & $-\frac{1}{2}$ & $-\frac{1}{2}$ & $0$ \\ \hline
\end{tabular}\label{TableLongSu2}

\smallskip

\centerline{\begin{tabular}{|c||c|c|c|c|}
  \hline $~$ & $e_1$ & $e_2$ & $e_3$ & $e_4$ \\
  \hline
  $\mathbb{L}^1_1$ & $\frac{1}{2}$ & $-\frac{1}{2}$ & $0$ & $0$ \\
  $\mathbb{R}^3_3$ & $0$ & $0$ & $\frac{1}{2}$ & $-\frac{1}{2}$ \\ \hline
\end{tabular}\label{TableShortSu2}}

\bigskip

From these tables it is now
easy to read off the blocks corresponding to weights $(a;b)$.
Explicitly, we find four one-dimensional blocks
\begin{align}
& V_{(\frac{3}{2};0)} = \{ f_6 \otimes e_1 \},& &
V_{(-\frac{3}{2};0)} = \{ f_8 \otimes e_2 \}, &\\
& V_{(0;\frac{3}{2})} = \{ f_9 \otimes e_3 \}, & &
V_{(0;-\frac{3}{2})} = \{ f_{11} \otimes e_4 \},
\end{align}
corresponding to vectors that have maximum/minumum weight. These
subspaces are related in the following way:
$(\Delta\mathbb{L}^1_2)^3 V_{(\frac{3}{2};0)} =
V_{(-\frac{3}{2};0)}$ and $(\Delta\mathbb{R}^3_4)^3
V_{(0;\frac{3}{2})} = V_{(0;-\frac{3}{2})}$.

\begin{figure}
  \centering
  \includegraphics[width=\textwidth]{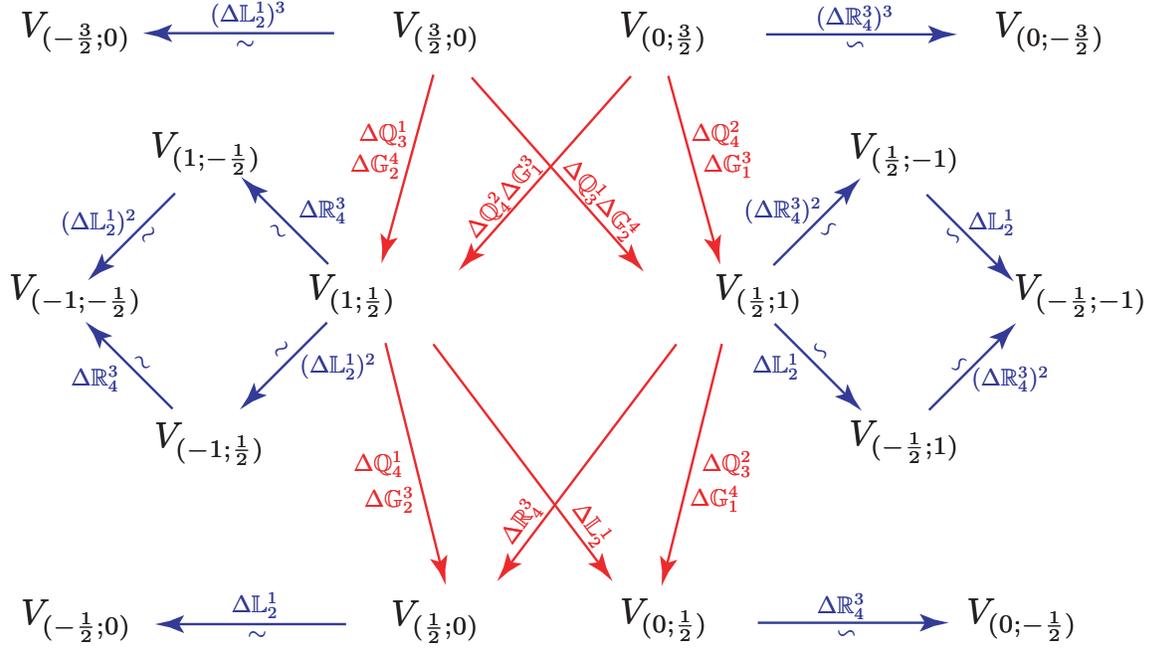}
  \caption{The relations between the different subspaces. The arrows with tildes denote isomorphic subspaces, which therefore have the same S-matrix block.}\label{Fig;Cases}
\end{figure}

Next, we have eight three-dimensional blocks
\begin{align}
 V_{(1;\frac{1}{2})} &= \{ f_2\otimes e_1, f_6
\otimes e_3, f_{12} \otimes e_1 \},&
V_{(-1;\frac{1}{2})} &= \{ f_3\otimes e_2, f_{8} \otimes e_3, f_{13}
\otimes e_2 \}, &\\
 V_{(1;-\frac{1}{2})} &= \{f_4 \otimes e_1, f_6 \otimes e_4, f_{14} \otimes e_1 \}, &
V_{(-1;-\frac{1}{2})} &= \{f_5 \otimes e_2, f_8 \otimes e_4, f_{15} \otimes e_2 \},
\end{align}
and
\begin{align}
 V_{(\frac{1}{2};1)} &= \{ f_2\otimes e_3, f_9
\otimes e_1, f_{12} \otimes e_3 \},&
V_{(\frac{1}{2};-1)} &= \{ f_4\otimes e_4, f_{11} \otimes e_1, f_{14}
\otimes e_4 \}, &\\
 V_{(-\frac{1}{2};1)} &= \{f_3 \otimes e_3, f_9 \otimes e_2, f_{13} \otimes e_3 \}, &
V_{(-\frac{1}{2};-1)} &= \{f_5 \otimes e_4, f_{11} \otimes e_2, f_{15} \otimes e_4 \},
\end{align}
Both sets of subspaces are again related {\it via} the
$\alg{su}(2) \oplus \alg{su}(2)$ generators as is indicated in
figure \ref{Fig;Cases}.

Finally, there are four nine-dimensional blocks
\begin{align}
 V_{(\frac{1}{2};0)} &= \{f_1\otimes e_1, f_{2} \otimes e_4, f_4
\otimes e_3, f_6 \otimes e_2, f_{7} \otimes e_1, f_{10}
\otimes e_1,
f_{12} \otimes e_4, f_{14} \otimes e_3, f_{16}
\otimes e_1 \},\nonumber\\
V_{(-\frac{1}{2};0)} &= \{f_1\otimes e_2, f_{3} \otimes e_4, f_5
\otimes e_3, f_7 \otimes e_2, f_{8} \otimes e_1, f_{10}
\otimes e_2,
f_{13} \otimes e_4, f_{15} \otimes e_3, f_{16}
\otimes e_2 \},\nonumber\\
V_{(0;\frac{1}{2})} &= \{f_1\otimes e_3, f_{2} \otimes e_2, f_9
\otimes e_1, f_7 \otimes e_3, f_{9} \otimes e_4, f_{10}
\otimes e_3,
f_{12} \otimes e_2, f_{13} \otimes e_1, f_{16}
\otimes e_3 \},\nonumber\\
V_{(0;-\frac{1}{2})} &= \{f_1\otimes e_4,f_{4} \otimes e_2, f_5 \otimes e_2, f_7 \otimes e_4, f_{10}
\otimes e_4, f_{11}\otimes e_3, f_{14} \otimes e_2, f_{15} \otimes e_1, f_{16}\otimes e_4 \}.
\end{align}

Because of the relations between the different subspaces one only
has to find the action of the S-matrix on the following subspaces
\begin{align}
 V_{(\frac{3}{2};0)},\quad
 V_{(0;\frac{3}{2})},\quad
 V_{(1;\frac{1}{2})},\quad
 V_{(\frac{1}{2};1)},\quad
 V_{(\frac{1}{2};0)},\quad
 V_{(0;\frac{1}{2})}.
\end{align}
Let us conveniently denote the vectors in these spaces by
\begin{align}
V_{(a;b)} = \{ |a;b\rangle_i \}_{i=1,\dim V_{(a;b)}}.
\end{align}
The final step consists in introducing a (opposite) coproduct basis
that allows for a quick derivation of the S-matrix. It turns out
that we must use as building blocks both
$|{\textstyle{\frac{3}{2}}};0\rangle$ and
$|0;{\textstyle{\frac{3}{2}}}\rangle$. We find for the
aforementioned three-dimensional subspaces
\begin{align}
 V_{(1;\frac{1}{2})} = \left\{ \Delta\mathbb{Q}^1_3|{\textstyle{\frac{3}{2}}};0\rangle,~
 \Delta\mathbb{G}^4_2|{\textstyle{\frac{3}{2}}};0\rangle,~
 \Delta\mathbb{Q}^2_4\Delta\mathbb{G}^3_1|0;{\textstyle{\frac{3}{2}}}\rangle
 \right\}, \\
 V_{(\frac{1}{2};1)} = \left\{ \Delta\mathbb{Q}^2_4|{0;\textstyle{\frac{3}{2}}}\rangle,~
 \Delta\mathbb{G}^3_1|{0;\textstyle{\frac{3}{2}}}\rangle,~
 \Delta\mathbb{Q}^1_3\Delta\mathbb{G}^4_2|{\textstyle{\frac{3}{2}}};0\rangle
 \right\}.
\end{align}
For the relevant nine-dimensional spaces we find
\begin{align}
 V_{(\frac{1}{2};0)} = \left\{
 \Delta\mathbb{R}^3_4|{\textstyle{\frac{1}{2}}};1\rangle_i,~
 \Delta\mathbb{Q}^1_4|1;{\textstyle{\frac{1}{2}}}\rangle_i,~
 \Delta\mathbb{G}^3_2|1;{\textstyle{\frac{1}{2}}}\rangle_i
 \right\}, \\
 V_{(0;\frac{1}{2})} = \left\{
 \Delta\mathbb{L}^1_2|1;{\textstyle{\frac{1}{2}}}\rangle_i ~
 \Delta\mathbb{Q}^1_4|{\textstyle{\frac{1}{2}}};1\rangle_i,~
 \Delta\mathbb{G}^3_2|{\textstyle{\frac{1}{2}}};1\rangle_i
 \right\},
\end{align}
where $i=1,2,3$.

\subsection{S-Matrix}

From the coproduct basis it is easily seen that the S-matrix will
be fixed upon specifying its action on
$|{\textstyle{\frac{3}{2}}};0\rangle$ and
$|0;{\textstyle{\frac{3}{2}}}\rangle$. Since these vectors both
form a one-dimensional block, they are mapped onto themselves by
the S-matrix. We will normalize the S-matrix as follows:
\begin{align}
\label{ics}
&\S|{\textstyle{\frac{3}{2}}};0\rangle =
|{\textstyle{\frac{3}{2}}};0\rangle,  &
\S|0;{\textstyle{\frac{3}{2}}}\rangle =
\mathscr{X}|0;{\textstyle{\frac{3}{2}}}\rangle.
\end{align}
Let us start considering the action of the S-matrix on the
three-dimensional block $V_{(1;\frac{1}{2})}$. We first define
\begin{align}
\label{qupm}
q_\pm \equiv \sqrt{q\pm 1}.
\end{align}
The basis transformation that relates the standard
basis to the coproduct basis and the opposite coproduct basis can
be written in terms of the following matrix:
\begin{align}
\Lambda_{3d, tot}(i,j,\kappa) = \begin{pmatrix}
 b_i q_- & d_i q_- & \kappa\left(b_j d_i - b_i d_j\right) q_+ \\
 a_j & c_j & - \kappa q_+ q_- \\
 - a_i q_+ & - c_i q_+ & \kappa \left(b_j c_i  -  a_i d_j\right) q_-
\end{pmatrix},
\end{align}
More precisely one finds that the basis transformations
$\Lambda_{3d}$ and $\Lambda_{3d}^{op}$ are given by
\begin{align}
&\Lambda_{3d} = \Lambda_{3d, tot}(1,2,1),&& \Lambda_{3d}^{op} =
\Lambda_{3d, tot}(3,4,\mathscr{X}).
\end{align}
We use the coefficients
(\ref{eqn;CoeffWithBraid1},\ref{eqn;CoeffWithBraid2}) that
explicitly include the braiding factors. By construction, the
action of the S-matrix is now given by
\begin{align}
\S| 1;{\textstyle{\frac{1}{2}}}\rangle_i =\sum_{j=1}^3
\mathscr{Y}_i^j | 1;{\textstyle{\frac{1}{2}}}\rangle_i,
\end{align}
with
\begin{align}
\mathscr{Y}= \Lambda^{op}_{3d}\Lambda^{-1}_{3d}.
\end{align}
The other three-dimensional space $V_{(\frac{1}{2};1)}$ has
transformation matrix
\begin{align}
\bar{\Lambda}_{3d,tot}(i,j,\kappa) = 
\begin{pmatrix}
 \kappa b_i q_+ & \kappa d_i q_+ & \left(a_j d_i - b_i c_j\right) q_- \\
 \kappa b_j & \kappa d_j & -q_+ q_- \\
 \kappa a_i q_- & \kappa c_i q_- & \left(a_i c_j - a_j c_i\right) q_+
\end{pmatrix} .
\end{align}
One again finds
\begin{align}
&\bar{\Lambda}_{3d} = \bar{\Lambda}_{3d, tot}(1,2,1),&&
\bar{\Lambda}_{3d}^{op} = \bar{\Lambda}_{3d,
tot}(3,4,\mathscr{X}).
\end{align}
This in turn leads to
\begin{align}
\S|{\textstyle{\frac{1}{2}}};1\rangle_i =\sum_{j=1}^3
\bar{\mathscr{Y}}_i^j | {\textstyle{\frac{1}{2}}};1\rangle_i,
\end{align}
with
\begin{align}
\left(\bar{\mathscr{Y}}^j_i\right)= \bar{\Lambda}^{op}_{3d}(\bar{\Lambda}_{3d})^{-1}.
\end{align}
The S-matrices in the other three-dimensional blocks are
also described by the above expressions. From Figure
\ref{Fig;Cases} we see that they are isomorphic via the
$\alg{su}(2)$ operators. They are related in a straightforward
way; the specified maps map basis vectors to basis vectors, e.g.
\begin{align}
\Delta R^3_4 |1;{\textstyle{\frac{1}{2}}}\rangle_i = |1;-{\textstyle{\frac{1}{2}}}\rangle_i.
\end{align}
This leads to
\begin{align}
&\S|\pm1;\pm{\textstyle{\frac{1}{2}}}\rangle_i =\sum_{j=1}^3
\mathscr{Y}_i^j | \pm1;\pm{\textstyle{\frac{1}{2}}}\rangle_i,&&
\S|\pm{\textstyle{\frac{1}{2}}};\pm1\rangle_i =\sum_{j=1}^3
\bar{\mathscr{Y}}_i^j |
\pm{\textstyle{\frac{1}{2}}};\pm1\rangle_i.
\end{align}
For the nine-dimensional blocks we again have two distinct cases.
We start with $V_{(\frac{1}{2};0)}$ and write
\begin{align}
\S|{\textstyle{\frac{1}{2}}};0\rangle_i =\sum_{j=1}^9
\mathscr{Z}_i^j |{\textstyle{\frac{1}{2}}};0\rangle_i.
\end{align}
Since we have expressed the coproduct basis in terms of the
three-dimensional subspaces, we find, in analogy with the previous
discussion,
\begin{align}
\left(\mathcal{Z}\right) = \Lambda_{9d}^{op}~ \diag(
\bar{\mathcal{Y}} , \mathcal{Y} , \mathcal{Y}) \Lambda_{9d}^{-1}.
\end{align}
The matrices $\Lambda_{9d}$ and $\Lambda^{op}_{9d}$ are
given by
\begin{align}
&\Lambda_{9d} = \left(\begin{smallmatrix}
 0 & 0 & 0 & b_1 \sqrt{q} & 0 & 0 & -d_1 \sqrt{q} & 0 & 0 \\
 1 & 0 & 0 & -a_2 & 0 & 0 & c_2 & 0 & 0 \\
 1 & 0 & 0 & 0 & b_1 q_- & 0 & 0 & -d_1 q_- & 0 \\
 0 & 0 & 0 & 0 & -b_2 & 0 & 0 & d_2 & 0 \\
 0 & 0 & 0 & -\frac{a_1 q_-}{\sqrt{2}} & 0 & \frac{b_1 q_+}{\sqrt{2}} & \frac{c_1 q_-}{\sqrt{2}} & 0 &
   -\frac{d_1 q_+}{\sqrt{2}} \\
 0 & \sqrt{2} & 0 & \frac{a_1 q_+}{\sqrt{2}} & 0 & \frac{b_1 q_-}{\sqrt{2}} & -\frac{c_1 q_+}{\sqrt{2}} & 0 &
   -\frac{d_1 q_-}{\sqrt{2}} \\
 0 & 0 & 1 & 0 & 0 & -a_2 & 0 & 0 & c_2 \\
 0 & 0 & 1 & 0 & -a_1 q_+ & 0 & 0 & c_1 q_+ & 0 \\
 0 & 0 & 0 & 0 & 0 & -a_1 \sqrt{q} & 0 & 0 & c_1 \sqrt{q}
\end{smallmatrix}\right),&& \Lambda_{9d}^{op} = \left.\Lambda_{9d}\right|_{(1\leftrightarrow3,2\leftrightarrow4)}.
\end{align}
On the other hand, we have
\begin{align}
\S| {0;\textstyle{\frac{1}{2}}}\rangle_i =\sum_{j=1}^9
\bar{\mathscr{Z}}_i^j |0;{\textstyle{\frac{1}{2}}}\rangle_i.
\end{align}
Here we find
\begin{align}
\left(\bar{\mathcal{Z}}\right) =
\bar{\Lambda}_{9d}^{op}~\diag(\mathcal{Y},
\bar{\mathcal{Y}},\bar{\mathcal{Y}}) (\bar{\Lambda}_{9d})^{-1}.
\end{align}
The matrices $\bar{\Lambda}_{9d}$ and $\bar{\Lambda}^{op}_{9d}$
are given by
\begin{align}
&\bar{\Lambda}_{9d} = \left(\begin{smallmatrix}
 0 & 0 & 0 & b_1 \sqrt{q} & 0 & 0 & -d_1 \sqrt{q} & 0 & 0 \\
 1 & 0 & 0 & b_2 & 0 & 0 & -d_2 & 0 & 0 \\
 1 & 0 & 0 & 0 & -b_1 q_+ & 0 & 0 & d_1 q_+ & 0 \\
 0 & \sqrt{2} & 0 & -\frac{a_1 q_-}{\sqrt{2}} & 0 & \frac{b_1 q_+}{\sqrt{2}} & \frac{c_1 q_-}{\sqrt{2}} & 0 &
   -\frac{d_1 q_+}{\sqrt{2}} \\
 0 & 0 & 0 & 0 & a_2 & 0 & 0 & -c_2 & 0 \\
 0 & 0 & 0 & \frac{a_1 q_+}{\sqrt{2}} & 0 & \frac{b_1 q_-}{\sqrt{2}} & -\frac{c_1 q_+}{\sqrt{2}} & 0 &
   -\frac{d_1 q_-}{\sqrt{2}} \\
 0 & 0 & 1 & 0 & 0 & b_2 & 0 & 0 & -d_2 \\
 0 & 0 & 1 & 0 & -a_1 q_- & 0 & 0 & c_1 q_- & 0 \\
 0 & 0 & 0 & 0 & 0 & -a_1 \sqrt{q} & 0 & 0 & c_1 \sqrt{q}
\end{smallmatrix}\right),&& \bar{\Lambda}_{9d}^{op} =
\left.\bar{\Lambda}_{9d}\right|_{(1\leftrightarrow3,2\leftrightarrow4)}.
\end{align}
Again one can use the $\alg{su}(2) \oplus \alg{su}(2)$ generators
to relate these two nine-dimensional S-matrices to the two
remaining ones. However, the relation is slightly less
straightforward. In particular we find
\begin{align}
\S|{0;-\textstyle{\frac{1}{2}}}\rangle_i &=\sum_{j=1}^9
(\mathscr{Z}^{\prime})_i^j |0;-{\textstyle{\frac{1}{2}}}\rangle_i, & (\mathscr{Z}^{\prime}) = L(\mathscr{Z})L^{-1},\\
\S|{-\textstyle{\frac{1}{2}}};0\rangle_i &=\sum_{j=1}^9
(\bar{\mathscr{Z}}^{\prime})_i^j |-{\textstyle{\frac{1}{2}}};0\rangle_i, & (\bar{\mathscr{Z}}^{\prime}) = R(\bar{\mathscr{Z}})R^{-1}.
\end{align}
where
\begin{align}
L&= \left(
\begin{smallmatrix}
 1 & 0 & 0 & 0 & 0 & 0 & 0 & 0 & 0 \\
 0 & 1 & 0 & 0 & 0 & 0 & 0 & 0 & 0 \\
 0 & 0 & 1 & 0 & 0 & 0 & 0 & 0 & 0 \\
 0 & 0 & 0 & \sqrt{2} & 1 & 0 & 0 & 0 & 0 \\
 0 & 0 & 0 & 0 & \sqrt{2} & 0 & 0 & 0 & 0 \\
 0 & 0 & 0 & 0 & 0 & 1 & 0 & 0 & 0 \\
 0 & 0 & 0 & 0 & 0 & 0 & 1 & 0 & 0 \\
 0 & 0 & 0 & 0 & 0 & 0 & 0 & 1 & 0 \\
 0 & 0 & 0 & 0 & 0 & 0 & 0 & 0 & 1
\end{smallmatrix}\right),
& R=\left(\begin{smallmatrix}
 1 & 0 & 0 & 0 & 0 & 0 & 0 & 0 & 0 \\
 0 & 1 & 0 & 0 & 0 & 0 & 0 & 0 & 0 \\
 0 & 0 & 1 & 0 & 0 & 0 & 0 & 0 & 0 \\
 0 & 0 & 0 & 1 & 0 & 0 & 0 & 0 & 0 \\
 0 & 0 & 0 & 0 & \sqrt{2} & 1 & 0 & 0 & 0 \\
 0 & 0 & 0 & 0 & 0 & \sqrt{2} & 0 & 0 & 0 \\
 0 & 0 & 0 & 0 & 0 & 0 & 1 & 0 & 0 \\
 0 & 0 & 0 & 0 & 0 & 0 & 0 & 1 & 0 \\
 0 & 0 & 0 & 0 & 0 & 0 & 0 & 0 & 1
\end{smallmatrix}\right).
\end{align}
It is easily checked that this S-matrix is symmetric and is indeed
invariant under the full centrally extended $\alg{su}(2|2)$
algebra. By construction, one can also see that the S-matrix is
automatically obtained in the factorized form $\S = F_{21} \,
F_{12}^{-1}$ (Drinfeld twist) \cite{twi}.

\subsection{Yang-Baxter Equation}

From the previous section we saw that invariance under the
symmetry algebra is not enough to fix the S-matrix completely. We
still have a free parameter $\mathscr{X}$. This parameter can be
fixed by imposing that the S-matrix solves the Yang-Baxter
equation:
\begin{align}
\S_{12}(P,p_2)\S_{13}(P,p_3)\S_{23}(p_2,p_3) = \S_{23}(p_2,p_3)\S_{13}(P,p_3)\S_{12}(P,p_2).
\end{align}
By considering the scattering processes
\begin{align}
&f_6\otimes e_1\otimes e_3 \rightarrow f_2\otimes e_1\otimes e_1,&
& f_6\otimes e_1\otimes e_2 \rightarrow f_6\otimes e_3\otimes e_4
\end{align}
we obtain two quadratic equations for $\mathscr{X}$ of the
form
\begin{align}
A + B \mathscr{X}(P,p_2) + C \mathscr{X}(P,p_3) + D\mathscr{X}(P,p_2)\mathscr{X}(P,p_3) =0,
\end{align}
where $A,B,C,D$ are functions of $P,p_2,p_3$. It is easily seen
that there are {\it two} different solutions to these equations.
This means that we find {\it two} S-matrices, and they are not
related by a similarity transformation. The solutions for
$\mathscr{X}$ appear however rather complicated and we refrain
from giving their explicit expressions. It can be checked
that {\it both} solutions for satisfy the following relations
\begin{description}
    \item[\it Unitarity:] $\qquad \, \, \, \, \, \, \, \, \, \, \, \, \S_{12}\S_{21}=\mathbbm{1}$.
    \item[\it Hermiticity:] $\, \, \, \, \, \, \qquad \S_{12}(z_L,z)\S_{12}(z_L^*,z^*)^{\dag} = \mathbbm{1}$.
    \item[\it CPT Invariance:] $\, \, \, \, \S_{12}=\S_{12}^t$.
    \item[\it Yang-Baxter:] $\, \, \, \qquad \S_{12}\S_{13}\S_{23}=\S_{23}\S_{13}\S_{12}$.
\end{description}
This completes our derivation of the S-matrices based on the
$\su(2|2)$ symmetry.

\section{Long representations via tensor product of short ones}\label{sect:short}

{\rm The scope of this section is to prove that the two S-matrices we have just derived, relying only on the Lie superalgebra symmetry and the Yang-Baxter equation, can both be obtained from the tensor product of two short evaluation representations of the Yangian. In principle from the previous analysis we could have found {\it more} solutions than those related to short evaluation representations, but we will show that this is not the case.}

Consider the tensor product of two short representations labelled
by momentum $(p_1,p_2)$,
\begin{align}
V(p_1)\otimes V(p_2).
\end{align}
This vector space naturally carries a representation of centrally
extended $\alg{su}(2|2)$ via the (opposite) coproduct, i.e. for
any generator $\mathbb{J}$ we have
\begin{align}
\label{delt}
\mathbb{J}_{V(p_1)\otimes V(p_2)} = \Delta\mathbb{J}.
\end{align}
It is easily seen by considering the central charges on this space
that we are dealing with a long representation. To be precise, we
find
\begin{align}
(2q)^2 = \Delta\mathbb{H}^2 - 4
\Delta\mathbb{C}\Delta\mathbb{C}^{\dag} = [E(p_1)+E(p_2)]^2 -
E(p_1+p_2)^2 + 1,
\end{align}
where the energy $E(p)$ is
\begin{align}
\label{sq} E(p)^2= 1+4g^2 \sin^2 \frac{p}{2}.
\end{align}
The momentum of the long representation is found to be
\begin{align}
P= p_1+p_2.
\end{align}
One has therefore
\begin{align}
V(p_1)\otimes V(p_2) \cong V(P,q)
\end{align}
with
\begin{align}\label{eqn;Paramters:LongviaShort}
&P=p_1+p_2, & ~&~ q =
\frac{E(p_1)+E(p_2)}{\sqrt{[E(p_1)+E(p_2)]^2}}\frac{\sqrt{[E(p_1)+E(p_2)]^2
- E(p_1+p_2)^2 + 1}}{2}.
\end{align}
The dispersion relation (\ref{sq}) has two branches, corresponding
to particles and anti-particles. Fixing momentum $p$ and a
choosing a branch specifies the fundamental representation
completely. Then, the tensor product of two such representations is
identified with a unique 16-dimensional long representation with
momentum $P$ and the central charge $q$ specified above.

Consider now the inverse problem, {\it i.e.} suppose we are given a
long representation $(P,q)$ and we want to factorize it into the
tensor product of two fundamental representations. It is
convenient to label the representation space corresponding to
particles as $V_+$ and the one corresponding to anti-particles as
$V_{-}$. Thus, the carrier space of the long representation can be
identified with one of the following four spaces:
\begin{enumerate}
\item[{ I}.] $V_+\otimes V_+$,
\item[{II}.] $V_+\otimes V_-$,
\item[{ III}.] $V_-\otimes V_+$,
\item[{ IV}.] $V_-\otimes V_-$.
\end{enumerate}
In the emerging solutions of the factorization problem the momenta
$p_1$ and $p_2$ can be ordered, and we always assume that the
ordering is such that  $p_1\prec p_2$\footnote{The details of the
ordering are irrelevant, since its only function is to choose a
unique representative between the couple $(p_1,p_2)$ and its permuted
couple $(p_2,p_1)$.}. Assuming for simplicity that $q$ is
real, we find that to a long representation $V(P,q)$ one can
associate two solutions of the  factorization problem. For
instance, for $q$ positive the two solutions are both associated
with the case I, or one of the solutions is from I and the second
is from II. Analogous situation takes place for $q$ negative.
Thus, any long representation can be written as a tensor product
of two {\it different} short representations. Actually, this
observation was reflected earlier in the fact that we found two
independent long-short S-matrices that solve the Yang-Baxter
equation.

Instead of particle momenta $p_i$ and $P$ one can use the
corresponding rapidity variables $u_i$ and $u$. The equation for
$q$ and the momentum conservation take the form
\begin{eqnarray}
\begin{aligned}\label{qrapid1} -
\Big[x\big(u_1-\frac{i}{g}\big)+x\big(u_1+\frac{i}{g}\big)+x\big(u_2-\frac{i}{g}\big)+x\big(u_2+\frac{i}{g}\big)
+\frac{2i}{g}\Big]^2+\\
+\frac{x\big(u+\frac{2i}{g}q\big)}{x\big(u-\frac{2i}{g}q\big)}
+\frac{x\big(u-\frac{2i}{g}q\big)}{x\big(u+\frac{2i}{g}q\big)}=4q^2+2\,
,
\end{aligned}
\end{eqnarray}
and
\begin{eqnarray}
\begin{aligned}\label{qrapid2}
\frac{x\big(u_1+\frac{i}{g}\big)}{x\big(u_1-\frac{i}{g}\big)}
\frac{x\big(u_2+\frac{i}{g}\big)}{x\big(u_2-\frac{i}{g}\big)}=
\frac{x\big(u+\frac{2i}{g}q\big)}{x\big(u-\frac{2i}{g}q\big)}\, ,
\end{aligned}
\end{eqnarray}
where $x(u)=\frac{u}{2}\Big(1+\sqrt{1-\frac{4}{u^2}}\Big)$ maps
the $u$-plane on the kinematic region of the string theory
\cite{Arutyunov:2007tc}. The energy of the long representation is
given by
\begin{eqnarray} E(P)=igx\Big(u-\frac{2i}{g}q\Big)-i g x\Big(u+\frac{2i}{g}q\Big)
-2q={\rm sign} (q)\, \sqrt{(2q)^2+4g^2\sin^2\frac{P}{2}}\,
\, .
\end{eqnarray}

In general, given $u_1$ and $u_2$,  the variable $q$ will appear
as a complicated function $q\equiv q(u_1,u_2,g)$. However, there
are two special cases, where $q$ is a constant independent of
$u_i$ and $g$. Indeed, for $u_1=u\pm\frac{i}{g}$ and
$u_2=u\mp\frac{i}{g}$ one gets $q=1$. Analogously, for
$u_1=-u\pm\frac{i}{g}$ and $u_2=-u\mp\frac{i}{g}$ one finds
$q=-1$. These values of $q$ correspond to the shortening
conditions, for which the long multiplet becomes reducible but
indecomposable. Imposing the ordering $u_1\prec u_2$ we get {\it
e.g. } for $q=1$ only one solution. This is an artifact of our
parametrization $x(u)$ in eqs.(\ref{qrapid1}) and (\ref{qrapid2}).
As is known, the $u$-plane covers through the map $x(u)$ only the
string region on the $z$-torus \cite{Arutyunov:2007tc}. To find
the other solution, one has to change the map $x(u)$ for the one
which covers the mirror regions on the $z$-torus. For $q=1$ both
solutions are from $V_+\otimes V_+$,  which is also the case for
$q$ close to one. However, when $q$ deviates from $q=1$
sufficiently enough, two solutions can occur in $V_+\otimes V_+$
and $V_+\otimes V_-$, respectively.

We further note that one can explicitly find the similarity
transformation that relates the long algebra generators to the
ones that arise from the coproduct. It is convenient to first
express the coefficients $a_L,b_L,c_L,d_L$ parameterizing long
representations via $a_1,$ etc. that describe the short
representations (again we use the coefficients that already include braiding factors, (\ref{eqn;CoeffWithBraid1},\ref{eqn;CoeffWithBraid2}))
\begin{align}
a_L &= \frac{a_1 d_1+a_2 d_2+q-1}{2 q d_L},  &  & b_L = \frac{d_L
\left(a_1 d_1 + a_2 d_2-q-1\right)}{c_1 d_1 + c_2 d_2}, & & c_L=
\frac{c_1 d_3 + c_2 d_2}{2q d_L}.
\end{align}
In terms of these coefficients we find that the algebra generators
are related via a similarity transformation
\begin{align}
\mathbb{J}_L = V_{\Delta}\Delta\mathbb{J}V^{-1}_{\Delta},
\end{align}
with
\begin{align}
\label{similcorto}
\setcounter{MaxMatrixCols}{16} V_{\Delta}=
\left(\begin{smallmatrix}
 0 & -v_1 & 0 & 0 & v_1 & 0 & 0 & 0 & 0 & 0 & 0 & -v_2 & 0 & 0 & v_2 & 0 \\
 0 & 0 & v_3 & 0 & 0 & 0 & 0 & 0 & v_4 & 0 & 0 & 0 & 0 & 0 & 0 & 0 \\
 0 & 0 & 0 & 0 & 0 & 0 & v_3 & 0 & 0 & v_4 & 0 & 0 & 0 & 0 & 0 & 0 \\
 0 & 0 & 0 & v_3 & 0 & 0 & 0 & 0 & 0 & 0 & 0 & 0 & v_4 & 0 & 0 & 0 \\
 0 & 0 & 0 & 0 & 0 & 0 & 0 & v_3 & 0 & 0 & 0 & 0 & 0 & v_4 & 0 & 0 \\
 \sqrt{2} v_5 & 0 & 0 & 0 & 0 & 0 & 0 & 0 & 0 & 0 & 0 & 0 & 0 & 0 & 0 & 0 \\
 0 & v_5 & 0 & 0 & v_5 & 0 & 0 & 0 & 0 & 0 & 0 & 0 & 0 & 0 & 0 & 0 \\
 0 & 0 & 0 & 0 & 0 & \sqrt{2} v_5 & 0 & 0 & 0 & 0 & 0 & 0 & 0 & 0 & 0 & 0 \\
 0 & 0 & 0 & 0 & 0 & 0 & 0 & 0 & 0 & 0 & \sqrt{2} v_6 & 0 & 0 & 0 & 0 & 0 \\
 0 & 0 & 0 & 0 & 0 & 0 & 0 & 0 & 0 & 0 & 0 & v_6 & 0 & 0 & v_6 & 0 \\
 0 & 0 & 0 & 0 & 0 & 0 & 0 & 0 & 0 & 0 & 0 & 0 & 0 & 0 & 0 & \sqrt{2} v_6 \\
 0 & 0 & v_7 & 0 & 0 & 0 & 0 & 0 & v_8 & 0 & 0 & 0 & 0 & 0 & 0 & 0 \\
 0 & 0 & 0 & 0 & 0 & 0 & v_7 & 0 & 0 & v_8 & 0 & 0 & 0 & 0 & 0 & 0 \\
 0 & 0 & 0 & v_7 & 0 & 0 & 0 & 0 & 0 & 0 & 0 & 0 & v_8 & 0 & 0 & 0 \\
 0 & 0 & 0 & 0 & 0 & 0 & 0 & v_7 & 0 & 0 & 0 & 0 & 0 & v_8 & 0 & 0 \\
 0 & -v_9 & 0 & 0 & v_9 & 0 & 0 & 0 & 0 & 0 & 0 & -v_{10} & 0 & 0 & v_{10} & 0
\end{smallmatrix}\right),
\end{align}
where the coefficients are given by
\begin{align}
v_1 &=  -\frac{\left(d_L a_1-b_L c_1\right) v_4}{\sqrt{q}} & v_2
&= \frac{\left(d_L b_2 - b_L d_2\right) v_4}{\sqrt{q}}, \nonumber\\
v_3 &= -\frac{\left(d_L a_1 - b_L c_1\right)v_4}{d_L a_2 - b_L
c_2}, &  v_5 &= -\frac{q_- \left(d_L a_1 - b_L c_1\right)
v_4}{\sqrt{2} \left(d_L a_2-b_L c_2\right)
   \left(d_L b_2-b_L d_2\right)}, \nonumber\\
v_6 &= -\frac{q_+v_4}{\sqrt{2} \left(d_L a_2 - b_L
   c_2\right)},&
v_7 & = \frac{q_+ q_- \left(d_L a_1 - b_L c_1\right) v_4}{2
\left(d_L a_2 - b_L c_2\right)^2
   \left(d_L b_2 - b_L d_2\right)}, \\
v_8 &= \frac{q_+ q_- v_4}{2 \left(d_L a_2 - b_L c_2\right)
\left(d_L b_2 - b_L d_2\right)}, & v_9 &= -\frac{q_+ q_-
\left(c_L a_1 - a_L c_1\right) v_4}{2 \sqrt{q} \left(d_L a_2 - b_L
c_2\right) \left(d_L b_2 - b_L d_2\right)},\nonumber\\
v_{10} &= \frac{q_+ q_- \left(c_L b_2 - a_L d_2\right) v_4}{2
\sqrt{q} \left(d_L a_2 - b_L c_2\right)
   \left(d_L b_2 - b_L d_2\right)}\nonumber.
\end{align}
The coproduct on three short representation is given by
$(\Delta\otimes\mathbbm{1})\Delta$. It is easily seen that
\begin{align}
\S_{13}\S_{23} (\Delta\otimes\mathbbm{1})\Delta\mathbb{J}
&= \S_{13}\S_{23} (\Delta\mathbb{J}\otimes\mathbb{U}^{[[\mathbb{J}]]}+\mathbbm{1}_L\otimes\mathbb{J}) \nonumber \\
&= \S_{13}\S_{23}
(\mathbb{J}\otimes\mathbb{U}^{[[\mathbb{J}]]}\otimes\mathbb{U}^{[[\mathbb{J}]]}+\mathbbm{1}\otimes\mathbb{J}\otimes\mathbb{U}^{[[\mathbb{J}]]}+\mathbbm{1}\otimes\mathbbm{1}\otimes\mathbb{J})
\nonumber \\
&=
(\mathbb{J}\otimes\mathbb{U}^{[[\mathbb{J}]]}\otimes\mathbbm{1}+\mathbbm{1}\otimes\mathbb{J}\otimes\mathbbm{1}+\mathbb{U}^{[[\mathbb{J}]]}\otimes\mathbb{U}^{[[\mathbb{J}]]}\otimes\mathbb{J})
\S_{13}\S_{23}\nonumber\\
&= (\Delta\mathbb{J}\otimes\mathbbm{1} +
{\mathbb{U}_L}^{[[\mathbb{J}]]}\otimes\mathbb{J})\S_{13}\S_{23}.
\end{align}
Thus we see that $\S_{13}\S_{23}$ intertwines the coproduct on the
tensor product of a long and a short representation. By the above
similarity transformation, this means that we can interpret $\S$
as being built up out of fundamental S-matrices.
\begin{align}
\S  = V_{\Delta}\otimes\mathbbm{1} \S_{13}\S_{23}
V^{-1}_{\Delta}\otimes\mathbbm{1}.
\end{align}
The two different choices of short representations that give rise
to the long representation then indeed gives two different
solutions for $\S$. They exactly coincide with the ones that are
found from the Yang-Baxter equation.

As we discussed in section \ref{sect:hy}, the fact that the
S-matrix in short representation possesses Yangian symmetries (in
evaluation representations) automatically induces, {\it via} the
above mentioned tensor product procedure, a Yangian representation
associated to the long representation. The generators are simply
given by
\begin{align}
\label{Ysh}
\widehat{\mathbb{J}}_{V(p_1)\otimes V(p_2)} = \Delta (\widehat{\mathbb{J}}).
\end{align}
$\Delta$ is projected into short $\otimes$ short Yangian
representations, the latter being characterized by the known
(`short') spectral parameters $u_1$ and $u_2$ (on the first and
second factor of the tensor product, respectively). These short
spectral parameters are linked to the parameters of the two
corresponding short representations as in (\ref{u}). When using
the formulas in appendices \ref{App;Yangian} and \ref{App:DrinII}
for the Yangian generators and their coproducts, taking into
account (\ref{delt}) and (\ref{Ysh}), one can check the perfect
consistency with all the relations in both Drinfeld's  first and
second realization, {\it and} one finds of course that the Yangian
in this representation {\rm admits an S-matrix} (in particular, the
higher central charges in Drinfeld's second realization are
co-commutative, and no contradition with the existence of an
S-matrix is found when acting on specific states, cf. section
\ref{sect:hy}). However, {\rm The Yangian representation obtained in this way is not isomorphic to the Yangian evaluation representation discussed in section \ref{sect:hy}. This is consistent with the fact that the evaluation representation of section \ref{sect:hy} does not admit an S-matrix, while the tensor product of two short representations does.}

\begin{figure}[h]
  \centering
  \includegraphics[width=\textwidth]{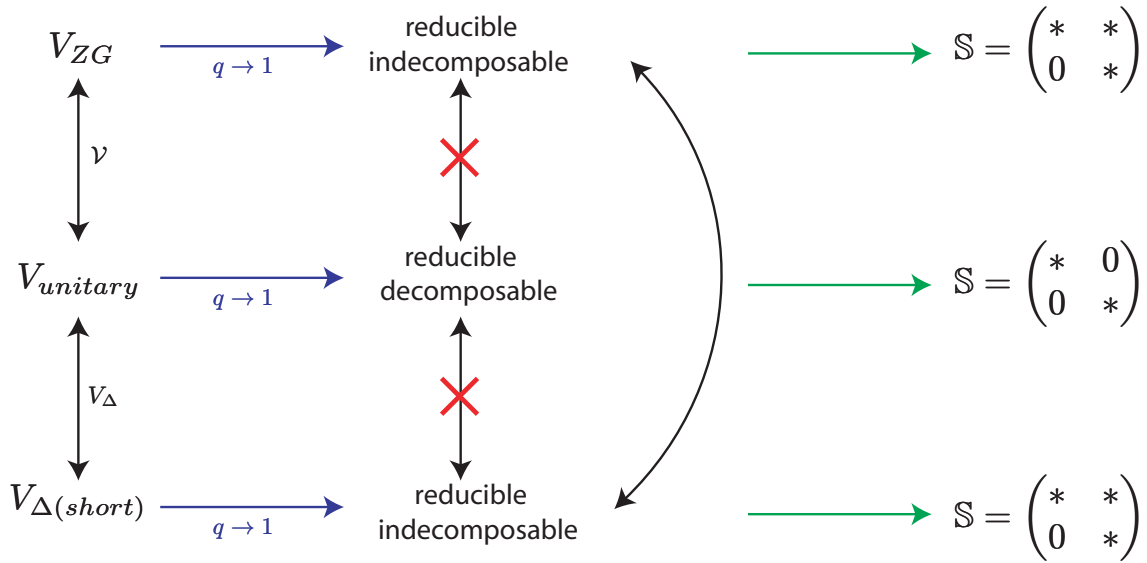}
  \caption{The various representations, their $q \to 1$ limits, and the block structure of the corresponding S-matrix in this limit. We denoted with $V_{ZG}$ the representation obtained from \cite{zhang-2005-46}. As a consequence of the upper-triangular structure, the bottom-right block of the limiting S-matrices satisfies the YBE by itself.}\label{Fig;dec}
\end{figure}

Let us {\rm add} one more remark on the situation corresponding to $q=1$. In this case, both the similarity transformation (\ref{simile})
that connects the Gould-Zhang representation \cite{zhang-2005-46} to the unitary one, and the one connecting the unitary representation to
a tensor product of short ones ({\it via} coproduct), namely (\ref{similcorto}), are singular. In fact, by sending $q \to 1$ in both the Gould-Zhang representation and in the tensor product of short ones, one gets a reducible but indecomposable representation, and the limit of the S-matrix is not block-diagonal in the subrepresentation and factor representation spaces. Instead, by sending $q \to 1$ in the unitary representation, one ends up into a decomposable representation (which is the only way it can be represented by hermitean matrices). The reducible components are the symmetric and antisymmetric bound-state representations, and the S-matrix trivally factorizes in the two spaces, each block becoming equal to the corresponding bound-state S-matrices. The relative unknown coefficient is not fixed, and this S-matrix trivially satisfies the Yang-Baxter equation blockwise, and has usual blockwise bound-state Yangian invariance in the evaluation representation. As we said, this unitary representation at $q=1$ is not
isomorphic to a tensor (or more precisely, to the co-) product of short fundamental representations, and
it turns out to furnish a way of extracting the subrepresentation and factor representation from the indecomposable one we are dealing with. We have summarized the situation for convenience in Figure 2.

\section*{Acknowledgements}
We are grateful to Sergey Frolov for many interesting discussions
and comments on the manuscript. We would like to thank Andrei
Babichenko, Romuald Janik, Nguyen Anh Ky and Adam Rej for their
comments. We would like to especially thank Niklas Beisert for
informing us about extra Serre relations arising in algebraic
extensions of the conventional Yangian. The work of G.~A. was
supported in part by the RFBR grant 08-01-00281-a, by the grant
NSh-795.2008.1, by NWO grant 047017015 and by the INTAS contract
03-51-6346.

\section{Appendices}

\subsection{Explicit Parameterization}\label{app:param}

We list in this appendix the generators of centrally extended
$\alg{su}(2|2)$ in the long representation. We only report
explicitly the simple roots for a distinguished Dynkin diagram,
the remainder of the algebra being generated {\it via} commutation
relations. We present the roots in a unitary representation. To
achieve this, we perform a similarity transformation on the
generator constructed directly from the oscillator basis of
\cite{zhang-2005-46}, in oder to obtain hermitean matrices. First,
the bosonic $\alg{su}(2)\oplus \su (2)$ roots are given by
\begin{align}
\setcounter{MaxMatrixCols}{16} \mathbb{L}^1_2 =
\left(\begin{smallmatrix}
0 & 0 & 0 & 0 & 0 & 0 & 0 & 0 & 0 & 0 & 0 & 0 & 0 & 0 & 0 & 0 \\
 0 & 0 & 0 & 0 & 0 & 0 & 0 & 0 & 0 & 0 & 0 & 0 & 0 & 0 & 0 & 0 \\
 0 & 1 & 0 & 0 & 0 & 0 & 0 & 0 & 0 & 0 & 0 & 0 & 0 & 0 & 0 & 0 \\
 0 & 0 & 0 & 0 & 0 & 0 & 0 & 0 & 0 & 0 & 0 & 0 & 0 & 0 & 0 & 0 \\
 0 & 0 & 0 & 1 & 0 & 0 & 0 & 0 & 0 & 0 & 0 & 0 & 0 & 0 & 0 & 0 \\
 0 & 0 & 0 & 0 & 0 & 0 & 0 & 0 & 0 & 0 & 0 & 0 & 0 & 0 & 0 & 0 \\
 0 & 0 & 0 & 0 & 0 & \sqrt{2} & 0 & 0 & 0 & 0 & 0 & 0 & 0 & 0 & 0 & 0 \\
 0 & 0 & 0 & 0 & 0 & 0 & \sqrt{2} & 0 & 0 & 0 & 0 & 0 & 0 & 0 & 0 & 0 \\
 0 & 0 & 0 & 0 & 0 & 0 & 0 & 0 & 0 & 0 & 0 & 0 & 0 & 0 & 0 & 0 \\
 0 & 0 & 0 & 0 & 0 & 0 & 0 & 0 & 0 & 0 & 0 & 0 & 0 & 0 & 0 & 0 \\
 0 & 0 & 0 & 0 & 0 & 0 & 0 & 0 & 0 & 0 & 0 & 0 & 0 & 0 & 0 & 0 \\
 0 & 0 & 0 & 0 & 0 & 0 & 0 & 0 & 0 & 0 & 0 & 0 & 0 & 0 & 0 & 0 \\
 0 & 0 & 0 & 0 & 0 & 0 & 0 & 0 & 0 & 0 & 0 & 1 & 0 & 0 & 0 & 0 \\
 0 & 0 & 0 & 0 & 0 & 0 & 0 & 0 & 0 & 0 & 0 & 0 & 0 & 0 & 0 & 0 \\
 0 & 0 & 0 & 0 & 0 & 0 & 0 & 0 & 0 & 0 & 0 & 0 & 0 & 1 & 0 & 0 \\
 0 & 0 & 0 & 0 & 0 & 0 & 0 & 0 & 0 & 0 & 0 & 0 & 0 & 0 & 0 & 0
\end{smallmatrix}\right),\quad
\mathbb{L}^2_1 = \left(\begin{smallmatrix}
 0 & 0 & 0 & 0 & 0 & 0 & 0 & 0 & 0 & 0 & 0 & 0 & 0 & 0 & 0 & 0 \\
 0 & 0 & 1 & 0 & 0 & 0 & 0 & 0 & 0 & 0 & 0 & 0 & 0 & 0 & 0 & 0 \\
 0 & 0 & 0 & 0 & 0 & 0 & 0 & 0 & 0 & 0 & 0 & 0 & 0 & 0 & 0 & 0 \\
 0 & 0 & 0 & 0 & 1 & 0 & 0 & 0 & 0 & 0 & 0 & 0 & 0 & 0 & 0 & 0 \\
 0 & 0 & 0 & 0 & 0 & 0 & 0 & 0 & 0 & 0 & 0 & 0 & 0 & 0 & 0 & 0 \\
 0 & 0 & 0 & 0 & 0 & 0 & \sqrt{2} & 0 & 0 & 0 & 0 & 0 & 0 & 0 & 0 & 0 \\
 0 & 0 & 0 & 0 & 0 & 0 & 0 & \sqrt{2} & 0 & 0 & 0 & 0 & 0 & 0 & 0 & 0 \\
 0 & 0 & 0 & 0 & 0 & 0 & 0 & 0 & 0 & 0 & 0 & 0 & 0 & 0 & 0 & 0 \\
 0 & 0 & 0 & 0 & 0 & 0 & 0 & 0 & 0 & 0 & 0 & 0 & 0 & 0 & 0 & 0 \\
 0 & 0 & 0 & 0 & 0 & 0 & 0 & 0 & 0 & 0 & 0 & 0 & 0 & 0 & 0 & 0 \\
 0 & 0 & 0 & 0 & 0 & 0 & 0 & 0 & 0 & 0 & 0 & 0 & 0 & 0 & 0 & 0 \\
 0 & 0 & 0 & 0 & 0 & 0 & 0 & 0 & 0 & 0 & 0 & 0 & 1 & 0 & 0 & 0 \\
 0 & 0 & 0 & 0 & 0 & 0 & 0 & 0 & 0 & 0 & 0 & 0 & 0 & 0 & 0 & 0 \\
 0 & 0 & 0 & 0 & 0 & 0 & 0 & 0 & 0 & 0 & 0 & 0 & 0 & 0 & 1 & 0 \\
 0 & 0 & 0 & 0 & 0 & 0 & 0 & 0 & 0 & 0 & 0 & 0 & 0 & 0 & 0 & 0 \\
 0 & 0 & 0 & 0 & 0 & 0 & 0 & 0 & 0 & 0 & 0 & 0 & 0 & 0 & 0 & 0
\end{smallmatrix}\right)
\end{align}
and
\begin{align}
\setcounter{MaxMatrixCols}{16} \mathbb{R}^3_4 =
\left(\begin{smallmatrix}
 0 & 0 & 0 & 0 & 0 & 0 & 0 & 0 & 0 & 0 & 0 & 0 & 0 & 0 & 0 & 0 \\
 0 & 0 & 0 & 0 & 0 & 0 & 0 & 0 & 0 & 0 & 0 & 0 & 0 & 0 & 0 & 0 \\
 0 & 0 & 0 & 0 & 0 & 0 & 0 & 0 & 0 & 0 & 0 & 0 & 0 & 0 & 0 & 0 \\
 0 & 1 & 0 & 0 & 0 & 0 & 0 & 0 & 0 & 0 & 0 & 0 & 0 & 0 & 0 & 0 \\
 0 & 0 & 1 & 0 & 0 & 0 & 0 & 0 & 0 & 0 & 0 & 0 & 0 & 0 & 0 & 0 \\
 0 & 0 & 0 & 0 & 0 & 0 & 0 & 0 & 0 & 0 & 0 & 0 & 0 & 0 & 0 & 0 \\
 0 & 0 & 0 & 0 & 0 & 0 & 0 & 0 & 0 & 0 & 0 & 0 & 0 & 0 & 0 & 0 \\
 0 & 0 & 0 & 0 & 0 & 0 & 0 & 0 & 0 & 0 & 0 & 0 & 0 & 0 & 0 & 0 \\
 0 & 0 & 0 & 0 & 0 & 0 & 0 & 0 & 0 & 0 & 0 & 0 & 0 & 0 & 0 & 0 \\
 0 & 0 & 0 & 0 & 0 & 0 & 0 & 0 & \sqrt{2} & 0 & 0 & 0 & 0 & 0 & 0 & 0 \\
 0 & 0 & 0 & 0 & 0 & 0 & 0 & 0 & 0 & \sqrt{2} & 0 & 0 & 0 & 0 & 0 & 0 \\
 0 & 0 & 0 & 0 & 0 & 0 & 0 & 0 & 0 & 0 & 0 & 0 & 0 & 0 & 0 & 0 \\
 0 & 0 & 0 & 0 & 0 & 0 & 0 & 0 & 0 & 0 & 0 & 0 & 0 & 0 & 0 & 0 \\
 0 & 0 & 0 & 0 & 0 & 0 & 0 & 0 & 0 & 0 & 0 & 1 & 0 & 0 & 0 & 0 \\
 0 & 0 & 0 & 0 & 0 & 0 & 0 & 0 & 0 & 0 & 0 & 0 & 1 & 0 & 0 & 0 \\
 0 & 0 & 0 & 0 & 0 & 0 & 0 & 0 & 0 & 0 & 0 & 0 & 0 & 0 & 0 & 0
\end{smallmatrix}\right),\quad
\mathbb{R}^4_3 = \left(\begin{smallmatrix}
 0 & 0 & 0 & 0 & 0 & 0 & 0 & 0 & 0 & 0 & 0 & 0 & 0 & 0 & 0 & 0 \\
 0 & 0 & 0 & 1 & 0 & 0 & 0 & 0 & 0 & 0 & 0 & 0 & 0 & 0 & 0 & 0 \\
 0 & 0 & 0 & 0 & 1 & 0 & 0 & 0 & 0 & 0 & 0 & 0 & 0 & 0 & 0 & 0 \\
 0 & 0 & 0 & 0 & 0 & 0 & 0 & 0 & 0 & 0 & 0 & 0 & 0 & 0 & 0 & 0 \\
 0 & 0 & 0 & 0 & 0 & 0 & 0 & 0 & 0 & 0 & 0 & 0 & 0 & 0 & 0 & 0 \\
 0 & 0 & 0 & 0 & 0 & 0 & 0 & 0 & 0 & 0 & 0 & 0 & 0 & 0 & 0 & 0 \\
 0 & 0 & 0 & 0 & 0 & 0 & 0 & 0 & 0 & 0 & 0 & 0 & 0 & 0 & 0 & 0 \\
 0 & 0 & 0 & 0 & 0 & 0 & 0 & 0 & 0 & 0 & 0 & 0 & 0 & 0 & 0 & 0 \\
 0 & 0 & 0 & 0 & 0 & 0 & 0 & 0 & 0 & \sqrt{2} & 0 & 0 & 0 & 0 & 0 & 0 \\
 0 & 0 & 0 & 0 & 0 & 0 & 0 & 0 & 0 & 0 & \sqrt{2} & 0 & 0 & 0 & 0 & 0 \\
 0 & 0 & 0 & 0 & 0 & 0 & 0 & 0 & 0 & 0 & 0 & 0 & 0 & 0 & 0 & 0 \\
 0 & 0 & 0 & 0 & 0 & 0 & 0 & 0 & 0 & 0 & 0 & 0 & 0 & 1 & 0 & 0 \\
 0 & 0 & 0 & 0 & 0 & 0 & 0 & 0 & 0 & 0 & 0 & 0 & 0 & 0 & 1 & 0 \\
 0 & 0 & 0 & 0 & 0 & 0 & 0 & 0 & 0 & 0 & 0 & 0 & 0 & 0 & 0 & 0 \\
 0 & 0 & 0 & 0 & 0 & 0 & 0 & 0 & 0 & 0 & 0 & 0 & 0 & 0 & 0 & 0 \\
 0 & 0 & 0 & 0 & 0 & 0 & 0 & 0 & 0 & 0 & 0 & 0 & 0 & 0 & 0 & 0
\end{smallmatrix}\right)~.
\end{align}
Next, we show two fermionic roots as an example:
\begin{align}
\setcounter{MaxMatrixCols}{16} \mathbb{Q}^1_3 =
\left(\begin{smallmatrix}
 \scriptscriptstyle 0 & \scriptscriptstyle 0 & \scriptscriptstyle 0 & \scriptscriptstyle -b \sqrt{q} & \scriptscriptstyle 0 & \scriptscriptstyle 0 & \scriptscriptstyle 0 & \scriptscriptstyle 0 & \scriptscriptstyle 0 & \scriptscriptstyle 0 & \scriptscriptstyle 0 & \scriptscriptstyle 0 & \scriptscriptstyle 0 & \scriptscriptstyle 0 & \scriptscriptstyle 0 & \scriptscriptstyle 0 \\
 \scriptscriptstyle 0 & \scriptscriptstyle 0 & \scriptscriptstyle 0 & \scriptscriptstyle 0 & \scriptscriptstyle 0 & \scriptscriptstyle b q_- & \scriptscriptstyle 0 & \scriptscriptstyle 0 & \scriptscriptstyle 0 & \scriptscriptstyle 0 & \scriptscriptstyle 0 & \scriptscriptstyle 0 & \scriptscriptstyle 0 & \scriptscriptstyle 0 & \scriptscriptstyle 0 & \scriptscriptstyle 0 \\
 \scriptscriptstyle a \sqrt{q} & \scriptscriptstyle 0 & \scriptscriptstyle 0 & \scriptscriptstyle 0 & \scriptscriptstyle 0 & \scriptscriptstyle 0 & \scriptstyle \frac{b q_-}{\sqrt{2}} & \scriptscriptstyle 0 & \scriptscriptstyle 0 & \scriptstyle \frac{b q_+}{\sqrt{2}} & \scriptscriptstyle 0 & \scriptscriptstyle 0 & \scriptscriptstyle 0 & \scriptscriptstyle 0 & \scriptscriptstyle 0 & \scriptscriptstyle 0 \\
 \scriptscriptstyle 0 & \scriptscriptstyle 0 & \scriptscriptstyle 0 & \scriptscriptstyle 0 & \scriptscriptstyle 0 & \scriptscriptstyle 0 & \scriptscriptstyle 0 & \scriptscriptstyle 0 & \scriptscriptstyle 0 & \scriptscriptstyle 0 & \scriptscriptstyle 0 & \scriptscriptstyle 0 & \scriptscriptstyle 0 & \scriptscriptstyle 0 & \scriptscriptstyle 0 & \scriptscriptstyle 0 \\
 \scriptscriptstyle 0 & \scriptscriptstyle 0 & \scriptscriptstyle 0 & \scriptscriptstyle 0 & \scriptscriptstyle 0 & \scriptscriptstyle 0 & \scriptscriptstyle 0 & \scriptscriptstyle 0 & \scriptscriptstyle 0 & \scriptscriptstyle 0 & \scriptscriptstyle b q_+ & \scriptscriptstyle 0 & \scriptscriptstyle 0 & \scriptscriptstyle 0 & \scriptscriptstyle 0 & \scriptscriptstyle 0 \\
 \scriptscriptstyle 0 & \scriptscriptstyle 0 & \scriptscriptstyle 0 & \scriptscriptstyle 0 & \scriptscriptstyle 0 & \scriptscriptstyle 0 & \scriptscriptstyle 0 & \scriptscriptstyle 0 & \scriptscriptstyle 0 & \scriptscriptstyle 0 & \scriptscriptstyle 0 & \scriptscriptstyle 0 & \scriptscriptstyle 0 & \scriptscriptstyle 0 & \scriptscriptstyle 0 & \scriptscriptstyle 0 \\
 \scriptscriptstyle 0 & \scriptscriptstyle 0 & \scriptscriptstyle 0 & \scriptstyle \frac{a q_-}{\sqrt{2}} & \scriptscriptstyle 0 & \scriptscriptstyle 0 & \scriptscriptstyle 0 & \scriptscriptstyle 0 & \scriptscriptstyle 0 & \scriptscriptstyle 0 & \scriptscriptstyle 0 & \scriptscriptstyle 0 & \scriptscriptstyle 0 & \scriptstyle -\frac{b q_+}{\sqrt{2}} & \scriptscriptstyle 0 & \scriptscriptstyle 0 \\
 \scriptscriptstyle 0 & \scriptscriptstyle 0 & \scriptscriptstyle 0 & \scriptscriptstyle 0 & \scriptscriptstyle a q_- & \scriptscriptstyle 0 & \scriptscriptstyle 0 & \scriptscriptstyle 0 & \scriptscriptstyle 0 & \scriptscriptstyle 0 & \scriptscriptstyle 0 & \scriptscriptstyle 0 & \scriptscriptstyle 0 & \scriptscriptstyle 0 & \scriptscriptstyle -b q_+ & \scriptscriptstyle 0 \\
 \scriptscriptstyle 0 & \scriptscriptstyle a q_+ & \scriptscriptstyle 0 & \scriptscriptstyle 0 & \scriptscriptstyle 0 & \scriptscriptstyle 0 & \scriptscriptstyle 0 & \scriptscriptstyle 0 & \scriptscriptstyle 0 & \scriptscriptstyle 0 & \scriptscriptstyle 0 & \scriptscriptstyle b q_- & \scriptscriptstyle 0 & \scriptscriptstyle 0 & \scriptscriptstyle 0 & \scriptscriptstyle 0 \\
 \scriptscriptstyle 0 & \scriptscriptstyle 0 & \scriptscriptstyle 0 & \scriptstyle \frac{a q_+}{\sqrt{2}} & \scriptscriptstyle 0 & \scriptscriptstyle 0 & \scriptscriptstyle 0 & \scriptscriptstyle 0 & \scriptscriptstyle 0 & \scriptscriptstyle 0 & \scriptscriptstyle 0 & \scriptscriptstyle 0 & \scriptscriptstyle 0 & \scriptstyle \frac{b q_-}{\sqrt{2}} & \scriptscriptstyle 0 & \scriptscriptstyle 0 \\
 \scriptscriptstyle 0 & \scriptscriptstyle 0 & \scriptscriptstyle 0 & \scriptscriptstyle 0 & \scriptscriptstyle 0 & \scriptscriptstyle 0 & \scriptscriptstyle 0 & \scriptscriptstyle 0 & \scriptscriptstyle 0 & \scriptscriptstyle 0 & \scriptscriptstyle 0 & \scriptscriptstyle 0 & \scriptscriptstyle 0 & \scriptscriptstyle 0 & \scriptscriptstyle 0 & \scriptscriptstyle 0 \\
 \scriptscriptstyle 0 & \scriptscriptstyle 0 & \scriptscriptstyle 0 & \scriptscriptstyle 0 & \scriptscriptstyle 0 & \scriptscriptstyle -a q_+ & \scriptscriptstyle 0 & \scriptscriptstyle 0 & \scriptscriptstyle 0 & \scriptscriptstyle 0 & \scriptscriptstyle 0 & \scriptscriptstyle 0 & \scriptscriptstyle 0 & \scriptscriptstyle 0 & \scriptscriptstyle 0 & \scriptscriptstyle 0 \\
 \scriptscriptstyle 0 & \scriptscriptstyle 0 & \scriptscriptstyle 0 & \scriptscriptstyle 0 & \scriptscriptstyle 0 & \scriptscriptstyle 0 & \scriptstyle -\frac{a q_+}{\sqrt{2}} & \scriptscriptstyle 0 & \scriptscriptstyle 0 & \scriptstyle \frac{a q_-}{\sqrt{2}} & \scriptscriptstyle 0 & \scriptscriptstyle 0 & \scriptscriptstyle 0 & \scriptscriptstyle 0 & \scriptscriptstyle 0 & \scriptscriptstyle -b \sqrt{q} \\
 \scriptscriptstyle 0 & \scriptscriptstyle 0 & \scriptscriptstyle 0 & \scriptscriptstyle 0 & \scriptscriptstyle 0 & \scriptscriptstyle 0 & \scriptscriptstyle 0 & \scriptscriptstyle 0 & \scriptscriptstyle 0 & \scriptscriptstyle 0 & \scriptscriptstyle 0 & \scriptscriptstyle 0 & \scriptscriptstyle 0 & \scriptscriptstyle 0 & \scriptscriptstyle 0 & \scriptscriptstyle 0 \\
 \scriptscriptstyle0 & \scriptscriptstyle 0 & \scriptscriptstyle 0 & \scriptscriptstyle 0 & \scriptscriptstyle 0 & \scriptscriptstyle 0 & \scriptscriptstyle 0 & \scriptscriptstyle 0 & \scriptscriptstyle 0 & \scriptscriptstyle 0 & \scriptscriptstyle a q_- & \scriptscriptstyle 0 & \scriptscriptstyle 0 & \scriptscriptstyle 0 & \scriptscriptstyle 0 & \scriptscriptstyle 0 \\
 \scriptscriptstyle 0 & \scriptscriptstyle 0 & \scriptscriptstyle 0 & \scriptscriptstyle 0 & \scriptscriptstyle 0 & \scriptscriptstyle 0 & \scriptscriptstyle 0 & \scriptscriptstyle 0 & \scriptscriptstyle 0 & \scriptscriptstyle 0 & \scriptscriptstyle 0 & \scriptscriptstyle 0 & \scriptscriptstyle 0 & \scriptstyle a \sqrt{q} & \scriptscriptstyle 0 & \scriptscriptstyle 0
\end{smallmatrix}\right)~,
\end{align}

\begin{align}
\setcounter{MaxMatrixCols}{16} \mathbb{G}^4_2 =
\left(\begin{smallmatrix}
\scriptscriptstyle 0 & \scriptscriptstyle 0 & \scriptscriptstyle 0 & \scriptscriptstyle -d \sqrt{q} & \scriptscriptstyle 0 & \scriptscriptstyle 0 & \scriptscriptstyle 0 & \scriptscriptstyle 0 & \scriptscriptstyle 0 & \scriptscriptstyle 0 & \scriptscriptstyle 0 & \scriptscriptstyle 0 & \scriptscriptstyle 0 & \scriptscriptstyle 0 & \scriptscriptstyle 0 & \scriptscriptstyle 0 \\
\scriptscriptstyle 0 & \scriptscriptstyle 0 & \scriptscriptstyle 0 & \scriptscriptstyle 0 & \scriptscriptstyle 0 & \scriptscriptstyle d q_- & \scriptscriptstyle 0 & \scriptscriptstyle 0 & \scriptscriptstyle 0 & \scriptscriptstyle 0 & \scriptscriptstyle 0 & \scriptscriptstyle 0 & \scriptscriptstyle 0 & \scriptscriptstyle 0 & \scriptscriptstyle 0 & \scriptscriptstyle 0 \\
\scriptscriptstyle c \sqrt{q} & \scriptscriptstyle 0 & \scriptscriptstyle 0 & \scriptscriptstyle 0 & \scriptscriptstyle 0 & \scriptscriptstyle 0 & \scriptscriptstyle \frac{d q_-}{\sqrt{2}} & \scriptscriptstyle 0 & \scriptscriptstyle 0 & \scriptscriptstyle \frac{d q_+}{\sqrt{2}} & \scriptscriptstyle 0 & \scriptscriptstyle 0 & \scriptscriptstyle 0 & \scriptscriptstyle 0 & \scriptscriptstyle 0 & \scriptscriptstyle 0 \\
\scriptscriptstyle 0 & \scriptscriptstyle 0 & \scriptscriptstyle 0 & \scriptscriptstyle 0 & \scriptscriptstyle 0 & \scriptscriptstyle 0 & \scriptscriptstyle 0 & \scriptscriptstyle 0 & \scriptscriptstyle 0 & \scriptscriptstyle 0 & \scriptscriptstyle 0 & \scriptscriptstyle 0 & \scriptscriptstyle 0 & \scriptscriptstyle 0 & \scriptscriptstyle 0 & \scriptscriptstyle 0 \\
\scriptscriptstyle 0 & \scriptscriptstyle 0 & \scriptscriptstyle 0 & \scriptscriptstyle 0 & \scriptscriptstyle 0 & \scriptscriptstyle 0 & \scriptscriptstyle 0 & \scriptscriptstyle 0 & \scriptscriptstyle 0 & \scriptscriptstyle 0 & \scriptscriptstyle d q_+ & \scriptscriptstyle 0 & \scriptscriptstyle 0 & \scriptscriptstyle 0 & \scriptscriptstyle 0 & \scriptscriptstyle 0 \\
\scriptscriptstyle 0 & \scriptscriptstyle 0 & \scriptscriptstyle 0 & \scriptscriptstyle 0 & \scriptscriptstyle 0 & \scriptscriptstyle 0 & \scriptscriptstyle 0 & \scriptscriptstyle 0 & \scriptscriptstyle 0 & \scriptscriptstyle 0 & \scriptscriptstyle 0 & \scriptscriptstyle 0 & \scriptscriptstyle 0 & \scriptscriptstyle 0 & \scriptscriptstyle 0 & \scriptscriptstyle 0 \\
\scriptscriptstyle 0 & \scriptscriptstyle 0 & \scriptscriptstyle 0 & \scriptscriptstyle \frac{c q_-}{\sqrt{2}} & \scriptscriptstyle 0 & \scriptscriptstyle 0 & \scriptscriptstyle 0 & \scriptscriptstyle 0 & \scriptscriptstyle 0 & \scriptscriptstyle 0 & \scriptscriptstyle 0 & \scriptscriptstyle 0 & \scriptscriptstyle 0 & \scriptscriptstyle -\frac{d q_+}{\sqrt{2}} & \scriptscriptstyle 0 & \scriptscriptstyle 0 \\
\scriptscriptstyle 0 & \scriptscriptstyle 0 & \scriptscriptstyle 0 & \scriptscriptstyle 0 & \scriptscriptstyle c q_- & \scriptscriptstyle 0 & \scriptscriptstyle 0 & \scriptscriptstyle 0 & \scriptscriptstyle 0 & \scriptscriptstyle 0 & \scriptscriptstyle 0 & \scriptscriptstyle 0 & \scriptscriptstyle 0 & \scriptscriptstyle 0 & \scriptscriptstyle -d q_+ & \scriptscriptstyle 0 \\
\scriptscriptstyle 0 & \scriptscriptstyle c q_+ & \scriptscriptstyle 0 & \scriptscriptstyle 0 & \scriptscriptstyle 0 & \scriptscriptstyle 0 & \scriptscriptstyle 0 & \scriptscriptstyle 0 & \scriptscriptstyle 0 & \scriptscriptstyle 0 & \scriptscriptstyle 0 & \scriptscriptstyle d q_- & \scriptscriptstyle 0 & \scriptscriptstyle 0 & \scriptscriptstyle 0 & \scriptscriptstyle 0 \\
\scriptscriptstyle 0 & \scriptscriptstyle 0 & \scriptscriptstyle 0 & \scriptscriptstyle \frac{c q_+}{\sqrt{2}} & \scriptscriptstyle 0 & \scriptscriptstyle 0 & \scriptscriptstyle 0 & \scriptscriptstyle 0 & \scriptscriptstyle 0 & \scriptscriptstyle 0 & \scriptscriptstyle 0 & \scriptscriptstyle 0 & \scriptscriptstyle 0 & \scriptscriptstyle \frac{d q_-}{\sqrt{2}} & \scriptscriptstyle 0 & \scriptscriptstyle 0 \\
\scriptscriptstyle 0 & \scriptscriptstyle 0 & \scriptscriptstyle 0 & \scriptscriptstyle 0 & \scriptscriptstyle 0 & \scriptscriptstyle 0 & \scriptscriptstyle 0 & \scriptscriptstyle 0 & \scriptscriptstyle 0 & \scriptscriptstyle 0 & \scriptscriptstyle 0 & \scriptscriptstyle 0 & \scriptscriptstyle 0 & \scriptscriptstyle 0 & \scriptscriptstyle 0 & \scriptscriptstyle 0 \\
\scriptscriptstyle 0 & \scriptscriptstyle 0 & \scriptscriptstyle 0 & \scriptscriptstyle 0 & \scriptscriptstyle 0 & \scriptscriptstyle -c q_+ & \scriptscriptstyle 0 & \scriptscriptstyle 0 & \scriptscriptstyle 0 & \scriptscriptstyle 0 & \scriptscriptstyle 0 & \scriptscriptstyle 0 & \scriptscriptstyle 0 & \scriptscriptstyle 0 & \scriptscriptstyle 0 & \scriptscriptstyle 0 \\
\scriptscriptstyle 0 & \scriptscriptstyle 0 & \scriptscriptstyle 0 & \scriptscriptstyle 0 & \scriptscriptstyle 0 & \scriptscriptstyle 0 & \scriptscriptstyle -\frac{c q_+}{\sqrt{2}} & \scriptscriptstyle 0 & \scriptscriptstyle 0 & \scriptscriptstyle \frac{c q_-}{\sqrt{2}} & \scriptscriptstyle 0 & \scriptscriptstyle 0 & \scriptscriptstyle 0 & \scriptscriptstyle 0 & \scriptscriptstyle 0 & \scriptscriptstyle -d \sqrt{q} \\
\scriptscriptstyle 0 & \scriptscriptstyle 0 & \scriptscriptstyle 0 & \scriptscriptstyle 0 & \scriptscriptstyle 0 & \scriptscriptstyle 0 & \scriptscriptstyle 0 & \scriptscriptstyle 0 & \scriptscriptstyle 0 & \scriptscriptstyle 0 & \scriptscriptstyle 0 & \scriptscriptstyle 0 & \scriptscriptstyle 0 & \scriptscriptstyle 0 & \scriptscriptstyle 0 & \scriptscriptstyle 0 \\
\scriptscriptstyle 0 & \scriptscriptstyle 0 & \scriptscriptstyle 0 & \scriptscriptstyle 0 & \scriptscriptstyle 0 & \scriptscriptstyle 0 & \scriptscriptstyle 0 & \scriptscriptstyle 0 & \scriptscriptstyle 0 & \scriptscriptstyle 0 & \scriptscriptstyle c q_- & \scriptscriptstyle 0 & \scriptscriptstyle 0 & \scriptscriptstyle 0 & \scriptscriptstyle 0 & \scriptscriptstyle 0 \\
\scriptscriptstyle 0 & \scriptscriptstyle 0 & \scriptscriptstyle 0 & \scriptscriptstyle 0 & \scriptscriptstyle 0 & \scriptscriptstyle 0 & \scriptscriptstyle 0 & \scriptscriptstyle 0 & \scriptscriptstyle 0 & \scriptscriptstyle 0 & \scriptscriptstyle 0 & \scriptscriptstyle 0 & \scriptscriptstyle 0 & \scriptscriptstyle c \sqrt{q} & \scriptscriptstyle 0 & \scriptscriptstyle 0
\end{smallmatrix}\right)~.
\end{align}
Finally, the similarity transformation that relates the unitary
representation to the one from \cite{zhang-2005-46} is given by
\begin{align}
\label{simile} \mathcal{V} =
\mathrm{diag}(\sqrt{q^3-q},q_+q_-,q_+q_-,q_+q_-,q_+q_-,2q_+,\sqrt{2}q_+,2q_+,2q_-,\sqrt{2}q_-,2q_-,1,1,1,1,\frac{1}{\sqrt{q}})
\end{align}
where $q_\pm$ have been defined in (\ref{qupm}).
We notice that this transformation is singular for $q^2=1$, where
the representation becomes reducible but indecomposable.

\subsection{Yangians and Coproducts: Drinfeld's first
realization}\label{App;Yangian}

The double Yangian \cite{Drin} $DY(\mathfrak{g})$ of a (simple) Lie algebra
$\mathfrak{g}$ is a deformation of the universal enveloping
algebra $U(\mathfrak{g}[u,u^{-1}])$ of the loop algebra
$\mathfrak{g}[u,u^{-1}]$. The Yangian is obtained by adding to the Lie algebra a set of partner
generators $\widehat{\mathbb{J}}^{A}_{n},\ n\in\mathbb{Z}$ satisfying the
commutation relations
\begin{eqnarray}
\ [\mathbb{J}^{A},\widehat{\mathbb{J}}^{B}  ]  = F^{AB}_{C} \,
\widehat{\mathbb{J}}^{C},
\end{eqnarray}
where $F^{AB}_{C}$ are the structure constants of $\mathfrak{g}$.
The centrally-extended $\alg{su}(2|2)$ Yangian has the following coproduct\footnote{Here, $\hbar = \frac{1}{g}$, and it can be reabsorbed in the definition of the algebra generators, which is the convention we use in the paper. We display $\hbar$ in this particular formula just to show how the `tail' of the Yangian coproduct organizes itself. The terms we omit from (\ref{eqn;CoProdYang}) are also completely determined by the knowledge of the level zero and level one coproducts, in a recursive fashion.}
\cite{Gomez:2006va,Plefka:2006ze,Beisert:2007ds}:
\begin{eqnarray}\label{eqn;CoProdYang}
\Delta (\, \widehat{\mathbb{J}}^{A}_n)  &=& \widehat{\mathbb{J}}^{A}_{n}\otimes
\mathbbm{1} + \mathbb{U}^{[[A]]}\otimes \widehat{\mathbb{J}}^{A}_{n} +
\hbar \sum_{m=0}^{n-1} F_{BC}^{A} \,
\mathbb{J}^{B}_{n-1-m}\mathbb{U}^{[[C]]}\otimes \mathbb{J}^{C}_{m} +\mathcal{O}(\hbar^{2}),
\end{eqnarray}
where $\mathbb{U}$, $[[A]]$ are given in section \ref{sect:hy}.

The evaluation representation we have been discussing in section \ref{sect:hy} is obtained as $\widehat{\mathbb{J}}^{A} =
u \, \mathbb{J}^{A}$ \cite{Beisert:2007ds}. In this representation the
coproduct structure is fixed in terms of the coproducts of
$\mathbb{J},\widehat{\mathbb{J}}$. As in the case of short representations, it is possible to absorb the factors arising due to the presence of $\mathbb{U}$ into a non-local redefinition of the representation labels
$a_i,b_i,c_i,d_i$. We will give here the formulas for the Yangian coproducts and for these redefined labels.
\begin{eqnarray}\label{eqn;CoProdYangDiff}
\Delta(\widehat{\mathbb{L}}^{a}_{\ b}) &=& \widehat{\mathbb{L}}^{\
a}_{1;b} + \widehat{\mathbb{L}}^{\ a}_{2;b} + \frac{1}{2}\mathbb{L}^{\
c}_{1;b}\mathbb{L}^{\ a}_{2;c}-\frac{1}{2} \mathbb{L}^{\
a}_{1;c}\mathbb{L}^{\ c}_{2;b}-\frac{1}{2} \mathbb{G}^{\
\gamma}_{1;b}\mathbb{Q}^{\ a}_{2;\gamma}-\frac{1}{2} \mathbb{Q}^{\
a}_{1;\gamma}\mathbb{G}^{\
\gamma}_{2;b}\nonumber\\
&&+\frac{1}{4}\delta^{a}_{b}\mathbb{G}^{\
\gamma}_{1;c}\mathbb{Q}^{\ c}_{2;\gamma}+\frac{1}{4}\delta^{a}_{b}
\mathbb{Q}^{\ c}_{1;\gamma}\mathbb{G}^{\
\gamma}_{2;c}~,\nonumber\\
\Delta(\widehat{\mathbb{R}}^{\alpha}_{\ \beta}) &=&
\widehat{\mathbb{R}}^{\ \alpha}_{1;\beta} + \widehat{\mathbb{R}}^{\
\alpha}_{2;\beta} - \frac{1}{2}\mathbb{R}^{\
\gamma}_{1;\beta}\mathbb{R}^{\ \alpha}_{2;\gamma}+\frac{1}{2}
\mathbb{R}^{\ \alpha}_{1;\gamma}\mathbb{R}^{\
\gamma}_{2;\beta}+\frac{1}{2} \mathbb{G}^{\
\alpha}_{1;c}\mathbb{Q}^{\ c}_{2;\beta}+\frac{1}{2} \mathbb{Q}^{\
c}_{1;\beta}\mathbb{G}^{\
\alpha}_{2;c}\nonumber\\
&&-\frac{1}{4}\delta^{\alpha}_{\beta}\mathbb{G}^{\
\gamma}_{1;c}\mathbb{Q}^{\
c}_{2;\gamma}-\frac{1}{4}\delta^{\alpha}_{\beta} \mathbb{Q}^{\
c}_{1;\gamma}\mathbb{G}^{\
\gamma}_{2;c}~,\nonumber\\
\Delta(\widehat{\mathbb{Q}}^{a}_{\ \beta}) &=& \widehat{\mathbb{Q}}^{\
a}_{1;\beta} + \widehat{\mathbb{Q}}^{\ a}_{2;\beta} -
\frac{1}{2}\mathbb{R}^{\ \gamma}_{1;\beta}\mathbb{Q}^{\
a}_{2;\gamma}+\frac{1}{2} \mathbb{Q}^{\ a}_{1;\gamma}\mathbb{R}^{\
\gamma}_{2;\beta} -\frac{1}{2} \mathbb{L}^{\ a}_{1;c}\mathbb{Q}^{\
c}_{2;\beta}+\frac{1}{2} \mathbb{Q}^{\
c}_{1;\beta}\mathbb{L}^{\ a}_{2;c}\nonumber\\
&&-\frac{1}{4}\mathbb{H}_{1}\mathbb{Q}^{\
a}_{2;\beta}+\frac{1}{4}\mathbb{Q}^{\ a}_{1;\beta}\mathbb{H}_{2} +
\frac{1}{2}\epsilon_{\beta\gamma}\epsilon^{ad}\mathbb{C}_{1}\mathbb{G}^{\
\gamma}_{2;d}-\frac{1}{2}\epsilon_{\beta\gamma}\epsilon^{ad}\mathbb{G}^{\
\gamma}_{1;d}\mathbb{C}_{2}~,\nonumber\\
\Delta(\widehat{\mathbb{G}}^{\alpha}_{\ b}) &=& \widehat{\mathbb{G}}^{\
\alpha}_{1;b} + \widehat{\mathbb{G}}^{\ \alpha}_{2;b} +
\frac{1}{2}\mathbb{L}^{\ c}_{1;b}\mathbb{G}^{\ \alpha}_{2;c}-
\frac{1}{2}\mathbb{G}^{\ \alpha}_{1;c}\mathbb{L}^{\ c}_{2;b}
+\frac{1}{2} \mathbb{R}^{\ \alpha}_{1;\gamma}\mathbb{G}^{\
\gamma}_{2;b} -\frac{1}{2} \mathbb{G}^{\
\gamma}_{1;b}\mathbb{R}^{\ \alpha}_{2;\gamma}\nonumber\\
&&+\frac{1}{4}\mathbb{H}_{1}\mathbb{G}^{\ \alpha}_{2;b}-
\frac{1}{4}\mathbb{G}^{\ \alpha}_{1;b}\mathbb{H}_{2} -
\frac{1}{2}\epsilon_{bc}\epsilon^{\alpha\gamma}\mathbb{C}^{\dag}_{1}\mathbb{Q}^{\
c}_{2;\gamma}
+\frac{1}{2}\epsilon_{bc}\epsilon^{\alpha\gamma}\mathbb{Q}^{\
c}_{1;\gamma}\mathbb{C}^{\dag}_{2}~,\nonumber\\
\Delta(\widehat{\mathbb{H}}) &=& \widehat{\mathbb{H}}_{1} +\widehat{\mathbb{H}}_{2}+ \mathbb{C}_{1}\mathbb{C}^{\dag}_{2}-\mathbb{C}^{\dag}_{1}\mathbb{C}_{2},\nonumber\\
\Delta(\widehat{\mathbb{C}}) &=& \widehat{\mathbb{C}}_{1}+\widehat{\mathbb{C}}_{2}-\frac{1}{2} \mathbb{H}_{1}\mathbb{C}_{2}+\frac{1}{2} \mathbb{C}_{1}\mathbb{H}_{2},\nonumber\\
\Delta(\widehat{\mathbb{C}}^{\dag}) &=&
\widehat{\mathbb{C}}^{\dag}_{1}+\widehat{\mathbb{C}}^{\dag}_{2}+\frac{1}{2}
\mathbb{H}_{1}\mathbb{C}^{\dag}_{2}-\frac{1}{2}
\mathbb{C}^{\dag}_{1}\mathbb{H}_{2}.
\end{eqnarray}
We have used in the above formulas the shorthand notation $\mathbb{J}_{1}\mathbb{Y}_{2} = \mathbb{J} \otimes \mathbb{Y}$. In case of long representation in space $1$ and short representation in space $2$ of the tensor product, the labels used in
$\Delta$ are given by:
\begin{eqnarray}\label{eqn;CoeffWithBraid1}
\begin{array}{lll}\
  a_{1} = \sqrt{\frac{g}{4q}}\eta_{1}, & ~ & b_{1} =
 -i e^{i p_{2}}\sqrt{\frac{g}{4q}}~
\frac{1}{\eta_{1}}\left(\frac{x_{1}^{+}}{x_{1}^{-}}-1\right), \\
  c_{1} = -e^{-i p_{2}}\sqrt{\frac{g}{4q}}\frac{\eta_{1}}{ x_{1}^{+}}, & ~ &
d_{1}=i\sqrt{\frac{g}{4q}}\frac{x_{1}^{+}}{\eta_{1}}\left(\frac{x_{1}^{-}}{x_{1}^{+}}-1\right),\\
\eta_{1} =
e^{i\frac{p_{1}}{4}}e^{i\frac{p_{2}}{2}}\sqrt{ix^{-}_{1}-ix^{+}_{1}},&~&~\\
~ &~& ~ \\
  a_{2} = \sqrt{\frac{g}{2}}\eta_{2}, & ~ & b_{2} = -i\sqrt{\frac{g}{2}}
\frac{1}{\eta_{2}}\left(\frac{x_{2}^{+}}{x_{2}^{-}}-1\right), \\
  c_{2} = -\sqrt{\frac{g}{2}}\frac{\eta_{2}}{x_{2}^{+}}, & ~ &
d_{2}=i\sqrt{\frac{g}{2}}\frac{x_{2}^{+}}{i\eta_{2}}\left(\frac{x_{2}^{-}}{x_{2}^{+}}-1\right),\\
\eta_{2} = e^{i\frac{p_{2}}{4}}\sqrt{ix^{-}_{2}-ix^{+}_{2}}.&~&~
\end{array}
\end{eqnarray}
Accordingly, the labels used in $\Delta^{op}$ are given by:
\begin{eqnarray}\label{eqn;CoeffWithBraid2}
\begin{array}{lll}
  a_{3} = \sqrt{\frac{g}{4q}}\eta^{op}_{1}, & ~ & b_{3} = -i\sqrt{\frac{g}{4q}}
\frac{1}{\eta^{op}_{1}}\left(\frac{x_{1}^{+}}{x_{1}^{-}}-1\right), \\
  c_{3} = -\sqrt{\frac{g}{4q}}\frac{\eta^{op}_{1}}{x_{1}^{+}}, & ~ &
d_{3}=i\sqrt{\frac{g}{4q}}\frac{x_{1}^{+}}{i\eta^{op}_{1}}\left(\frac{x_{1}^{-}}{x_{1}^{+}}-1\right),\\
\eta^{op}_{1} =
e^{i\frac{p_{1}}{4}}\sqrt{ix^{-}_{1}-ix^{+}_{1}},&~&~\\
 ~ &~& ~ \\
  a_{4} = \sqrt{\frac{g}{2}}\eta^{op}_{2}, & ~ &
b_{4} =
 -i e^{i p_{1}}\sqrt{\frac{g}{2}}~
\frac{1}{\eta^{op}_{2}}\left(\frac{x_{2}^{+}}{x_{2}^{-}}-1\right), \\
  c_{4} = -e^{-i p_{1}}\sqrt{\frac{g}{2}}\frac{\eta^{op}_{2}}{ x_{2}^{+}}, & ~ &
d_{4}=i\sqrt{\frac{g}{2}}\frac{x_{2}^{+}}{\eta^{op}_{2}}\left(\frac{x_{2}^{-}}{x_{2}^{+}}-1\right),\\
\eta^{op}_{2} =
e^{i\frac{p_{2}}{4}}e^{i\frac{p_{1}}{2}}\sqrt{ix^{-}_{2}-ix^{+}_{2}}.&~&~
\end{array}
\end{eqnarray}
The non-trivial braiding factors are all hidden in the parameters
of the four representations involved.

\subsection{Yangians and Coproducts: Drinfeld's second
realization}\label{App:DrinII}

The second realization of the Yangian \cite{Dsecond} is given in terms of
Chevalley-Serre type generators and relations. The formulas for the centrally-extended $\alg{su} (2|2)$ case have been given in \cite{Spill:2008tp}. They are expressed in terms of Cartan generators $\kappa_{i,m}$ and fermionic simple roots $\xi^\pm_{i,m}$, $i=1,2,3$, $m=0,1,2,\dots$, subject to the following relations:
\begin{align}
\label{relazionizero}
&[\kappa_{i,m},\kappa_{j,n}]=0,\quad [\kappa_{i,0},\xi^+_{j,m}]=a_{ij} \,\xi^+_{j,m},\nonumber\\
&[\kappa_{i,0},\xi^-_{j,m}]=- a_{ij} \,\xi^-_{j,m},\quad \{\xi^+_{i,m},\xi^-_{j,n}\}=\delta_{i,j}\, \kappa_{j,n+m},\nonumber\\
&[\kappa_{i,m+1},\xi^+_{j,n}]-[\kappa_{i,m},\xi^+_{j,n+1}] = \frac{1}{2} a_{ij} \{\kappa_{i,m},\xi^+_{j,n}\},\nonumber\\
&[\kappa_{i,m+1},\xi^-_{j,n}]-[\kappa_{i,m},\xi^-_{j,n+1}] = - \frac{1}{2} a_{ij} \{\kappa_{i,m},\xi^-_{j,n}\},\nonumber\\
&\{\xi^+_{i,m+1},\xi^+_{j,n}\}-\{\xi^+_{i,m},\xi^+_{j,n+1}\} = \frac{1}{2} a_{ij} [\xi^+_{i,m},\xi^+_{j,n}],\nonumber\\
&\{\xi^-_{i,m+1},\xi^-_{j,n}\}-\{\xi^-_{i,m},\xi^-_{j,n+1}\} = - \frac{1}{2} a_{ij} [\xi^-_{i,m},\xi^-_{j,n}],
\end{align}
\begin{eqnarray}
\label{relazero}
&&i\neq j, \, \, \, \, \, n_{ij}=1+|a_{ij}|,\, \, \, \, \, Sym_{\{k\}} [\xi^+_{i,k_1},[\xi^+_{i,k_2},\dots \{\xi^+_{i,k_{n_{ij}}}, \xi^+_{j,l}\}\dots\}\}=0,\nonumber\\
&&i\neq j, \, \, \, \, \, n_{ij}=1+|a_{ij}|,\, \, \, \, \, Sym_{\{k\}} [\xi^-_{i,k_1},[\xi^-_{i,k_2},\dots \{\xi^-_{i,k_{n_{ij}}}, \xi^-_{j,l}\}\dots\}\}=0,\nonumber\\
&&\text{except for} \, \, \, \, \, \, \, \, \, \{\xi^+_{2,n},\xi^+_{3,m}\} = \fC_{n+m}, \qquad  \{\xi^-_{2,n},\xi^-_{3,m}\} = \fC^\dag_{n+m},
\end{eqnarray}
where the symmetric
Cartan matrix $a_{ij}$ has all zeroes except for $a_{12}=a_{21}=1$ and $a_{13}=a_{31}=-1$. We call the index $n$ of the generators in this realization the {\it level}. The Dynkin diagram corresponds to the following Chevalley-Serre basis, composed of Cartan generators $\gen{H}_i$, and
positive (negative) simple roots $\gen{E}^+_i$ ($\gen{E}^-_i$,
respectively) \cite{Spill:2008tp}
\begin{align}
\gen{E}_1^+ &= \fGG^4_2, \qquad &\gen{E}_1^- &= \fQ^2_4, \qquad &\gen{H}_1 & = -\fL_1^1 - \fR^3_3 + \frac{1}{2}\fHH,\\
\gen{E}_2^+ &= i\fQ^1_4, \qquad &\gen{E}_2^- &= i\fGG^4_1, \qquad &\gen{H}_2 & = -\fL_1^1 + \fR^3_3 - \frac{1}{2}\fHH,\\
\gen{E}_3^+ &= i\fQ^2_3, \qquad &\gen{E}_3^- &= i\fGG^3_2, \qquad &\gen{H}_3 & = \fL_1^1 - \fR^3_3 - \frac{1}{2}\fHH.
\end{align}
The isomorphism (Drinfeld's map) between the first and the second realization is given
as follows:
\begin{align}
\label{mappa1}
&\kappa_{i,0}=\gen{H}_i,\quad \xi^+_{i,0}=\gen{E}_i,\quad \xi^-_{i,0}=\gen{F}_i,\nonumber\\
&\kappa_{i,1}=\widehat{\gen{H}}_i-v_i,\quad \xi^+_{i,1}=\widehat{\gen{E}}_i-w_i,\quad \xi^-_{i,1}=\widehat{\gen{F}}_i-z_i,
\end{align}
where $\widehat{\gen{H}}_i, \widehat{\gen{E}}_i, \widehat{\gen{F}}_i$ are the Yangian partners of $\gen{H}_i, \gen{E}_i, \gen{F}_i$ in the first realization, and the special elements are given by
\begin{eqnarray}
v_1 &=& - \frac{1}{2} \kappa_{1,0}^2 + \frac{1}{4} \fR^4_3 \fR^3_4 +  \frac{1}{4} \fR^3_4 \fR^4_3 +  \frac{3}{4} \fL^2_1 \fL^1_2 - \frac{1}{4} \fL^1_2 \fL^2_1 - \frac{1}{4} \fQ^2_3 \fGG^3_2 - \frac{1}{4} \fQ^1_4 \fGG^4_1 - \frac{3}{4} \fGG^4_1 \fQ^1_4  \nonumber\\
&&+ \frac{1}{4} \fGG^3_2 \fQ^2_3 + \half \fC \fC^\dag ,\nonumber\\
v_2 &=& - \frac{1}{2} \kappa_{2,0}^2 - \fR^4_3 \fR^3_4 +  \frac{1}{2} \fR^3_4 \fR^4_3 + \frac{1}{2} \fL^1_2 \fL^2_1 + \fQ^1_3 \fGG^3_1 + \frac{1}{2} \fGG^3_1 \fQ^1_3  - \frac{1}{2} \fGG^4_2 \fQ^2_4 - \half \fC \fC^\dag ,\nonumber\\
v_3 &=& - \frac{1}{2} \kappa_{3,0}^2 + \frac{1}{2} \fR^4_3 \fR^3_4  - \frac{1}{2} \fL^2_1 \fL^1_2 + \frac{1}{2} \fGG^3_1 \fQ^1_3  + \frac{1}{2} \fGG^4_2 \fQ^2_4 - \half \fC \fC^\dag ,\nonumber \\
w_1 &=& - \frac{1}{4} (\xi^+_{1,0} \kappa_{1,0} + \kappa_{1,0} \xi^+_{1,0}) + \frac{3}{4} \fGG^4_1 \fL^1_2 -\frac{1}{4} \fL^1_2 \fGG^4_1 + \frac{1}{4} \fGG^3_2 \fR^4_3 + \frac{1}{4} \fR^4_3 \fGG^3_2 + \frac{1}{2} \fQ^1_3 \fC^\dag,\nonumber\\
w_2 &=& - \frac{1}{4} (\xi^+_{2,0} \kappa_{2,0} + \kappa_{2,0} \xi^+_{2,0}) + \frac{3 i}{4} \fQ^1_3 \fR^3_4 -\frac{i}{4} \fL^1_2 \fQ^2_4 - \frac{i}{4} \fQ^2_4 \fL^1_2 - \frac{i}{4} \fR^3_4 \fQ^1_3 - \frac{i}{2} \fGG^3_2 \fC,\nonumber\\
w_3 &=& - \frac{1}{4} (\xi^+_{3,0} \kappa_{3,0} + \kappa_{3,0} \xi^+_{3,0}) - \frac{i}{4} \fQ^1_3 \fL^2_1 +\frac{3 i}{4} \fL^2_1 \fQ^1_3 - \frac{i}{4} \fQ^2_4 \fR^4_3 - \frac{i}{4} \fR^4_3 \fQ^2_4 - \frac{i}{2} \fGG^4_1 \fC,\nonumber\\
z_1 &=& - \frac{1}{4} (\xi^-_{1,0} \kappa_{1,0} + \kappa_{1,0} \xi^-_{1,0}) - \frac{1}{4} \fQ^1_4 \fL^2_1 +\frac{3}{4} \fL^2_1 \fQ^1_4 + \frac{1}{4} \fQ^2_3 \fR^3_4 + \frac{1}{4} \fR^3_4 \fQ^2_3 + \frac{1}{2} \fGG^3_1 \fC,\nonumber\\
z_2 &=& - \frac{1}{4} (\xi^-_{2,0} \kappa_{2,0} + \kappa_{2,0} \xi^-_{2,0}) - \frac{i}{4} \fGG^3_1 \fR^4_3 +\frac{3 i}{4} \fR^4_3 \fGG^3_1 - \frac{i}{4} \fGG^4_2 \fL^2_1 - \frac{i}{4} \fL^2_1 \fGG^4_2 - \frac{i}{2} \fQ^2_3 \fC^\dag,\nonumber\\
z_3 &=& - \frac{1}{4} (\xi^-_{3,0} \kappa_{3,0} + \kappa_{3,0} \xi^-_{3,0}) - \frac{i}{4} \fGG^4_2 \fR^3_4 - \frac{i}{4} \fR^3_4 \fGG^4_2 + \frac{3 i}{4} \fGG^3_1 \fL^1_2 - \frac{i}{4} \fL^1_2 \fGG^3_1 - \frac{i}{2} \fQ^1_4 \fC^\dag.\nonumber
\end{eqnarray}
By knowing level-zero and level-one generators, one can recursively construct all higher-level generators by repeated use of the relations (\ref{relazionizero}). We have performed extensive checks of the consistency of the (long) representation we find after Drinfeld's map with all relations (\ref{relazionizero}). The explicit form of these generators is not particularly illuminating and we omit to report it here. The only interesting point is that it is not of a simple evaluation type, but rather more complicated. The Cartan generators at level one, for instance, are not represented by diagonal matrices, still with all relations (\ref{relazionizero}) being satisfied.

The above reported Drinfeld's map is  also used to derive the
Yangian coproducts in Drinfeld's second realization by knowing the
coproducts in Drinfeld's first realization (see previous section)
and using the homomorphism property $\Delta(a b) = \Delta(a)
\Delta(b)$. Same consistency we have found for coproducts and
other Hopf algebra structures.

\subsection{A remark on long representations and Hirota equations}\label{sec:Hirota}
The large $L$ asymptotic solution for the Y-system (see the
Introduction) is most conveniently written in terms of certain
transfer-matrices associated with the underlying symmetry group of
the model \cite{Kuniba:1993cn}. In the context of the string sigma
model the corresponding asymptotic solution was presented in
\cite{Gromov:2009tv}. In this solution the corresponding
Y-functions are re-expressed in terms of suitable T-functions
$T_{a,s}$. The latter must obey the so-called Hirota equations
\begin{eqnarray}
\label{Hiro}
T^+_{a,s} (u)\, T^-_{a,s} (u)\, = \, T_{a+1,s} (u)\, T_{a-1,s} (u)\, + \, T_{a,s+1} (u)\, T_{a,s-1} (u),
\end{eqnarray}
where $f^\pm (u) = f(u\pm \frac{i}{g})$. These equations are
formally solved by the Bazhanov-Reshetikhin (BR) determinant formula
\cite{Bazhanov:1989yk}
\begin{eqnarray}\nonumber
T_{a,s}(u) &= \det_{1\leq i,j\leq s} T_{a+i-j,1}(u
+\frac{i}{g}(s+1-i-j))=\\
&~~~~~~~~~~=\det_{1\leq i,j\leq a} T_{1,s+i-j}(u
+\frac{i}{g}(a+1-i-j))\, ,
\end{eqnarray}
which expresses all $T_{a,s}$ either in terms of $T_{1,s}$ or
$T_{a,1}$. In the large $L$-limit the T-function $T_{a,s}$ is
supposed to coincide with (the eigenvalues of) the transfer matrix
evaluated in the rectangular representation $(a,s)$ of the
centrally extended $\alg{sl}(2|2)$. For the case without central
extension, this fact has been proved in \cite{Kazakov:2007fy}.
Here we will be concerned with centrally-extended $\alg{sl}(2|2)$.
Rather than developing a general theory, we will construct
explicitly for one simple example the corresponding transfer
matrices, {\it i.e. without} appealing to the BR formula, and show
that the Hirota equations are indeed satisfied.

Our construction also allows one to better understand the role of
long representations giving rise to a generic transfer matrix
$T_{L}$. Namely, long representations for which the central charge
$q$ satisfies the shortening condition become reducible but
indecomposable. Using the relationship between long and short
representations, we show that the transfer matrix $T_{L}$ admit a
factorization into a tensor product of the transfer-matrices
corresponding to short representations. We carry out this
construction for our simplest 16-dimensional long representation
with $q=\pm 1$. For these values, the corresponding transfer
matrix $T_{L}(u)$ admits a factorisation
\begin{eqnarray}
\label{FactorLong} T_{L}^{16|1}(u)=T^{4|\frac{1}{2}}_{1,1}
\big(u+\frac{i}{g}\big)\, T^{4|\frac{1}{2}}_{1,1}
\big(u-\frac{i}{g}\big)\, .
\end{eqnarray} Obviously, this is the left hand side of (\ref{Hiro}).
Here we use the notation $T^{{\rm dim}|q}$ to indicate the
dimension and the charge $q$ of the corresponding representation.
To write down the right hand side, we recall that the Hirota
equations are invariant under a certain gauge symmetry. This
symmetry can be used to set $T_{0,s}=1$ for all $s$. Since in the
large $L$ limit $T_{a,0}=1$ for all $a$, the Hirota equation takes
the form\footnote{Notice that this can also be interpreted in the light of the discussion at the end of section 4. The transfer matrix is insensitive to a similarity transformation in the auxiliary space, therefore the unitary representation should really give the same result as the tensor (or co-) product of short ones. In fact, in the unitary representation the above splitting in the limit of $q \to 1$ is an obvious consequence of decomposability into the bound-state irreducible components.}
\begin{eqnarray}\label{simplestHiro}
T_{L}^{16|1}(u)=T^{4|\frac{1}{2}}_{1,1} \big(u+\frac{i}{g}\big)\,
T^{4|\frac{1}{2}}_{1,1} \big(u-\frac{i}{g}\big) =\, T_{2,1}^{8|1}
(u)\, + \, T_{1,2}^{8|1} (u).
\end{eqnarray}
Here $T_{2,1}$ and $T_{1,2}$ are the transfer matrices
corresponding to short 8-dimensional anti-symmetric and symmetric
representations, respectively. These are precisely those which
appear as the subrepresentation and the factor representation of
the 16-dimensional long multiplet with $q=1$. Obviously,
eq.(\ref{simplestHiro}) represents the fusion mechanism.

All transfer matrix eigenvalues $T_{1,s}$ has been obtained
with the help of the Algebraic Bethe Ansatz technique in \cite{Arutyunov:2009iq}.
Alternatively, they can be found with the help of the quantum
characteristic function \cite{Beisert:2006qh}. Here, we first need the
eigenvalues which correspond to the $\su(2)$ sector. They are
given by \begin{eqnarray}\label{TS}
T_{\su(2)}(u\,|\,\vec{v})&=&1+\prod_{i=1}^{K^{\rm{I}}}
\frac{(x^--x^-_i)(1-x^- x^+_i)}{(x^+-x^-_i)(1-x^+
x^+_i)}\frac{x^+}{x^-}\\
&&\hspace{-0.5cm} -2\sum_{k=0}^{s-1}\prod_{i=1}^{K^{\rm{I}}}
\frac{x^+-x^+_i}{x^+-x^-_i}\sqrt{\frac{x^-_i}{x^+_i}}
\left[1-\frac{\frac{2ik}{g}}{u-v_i+\frac{i}{g}(s-1)}\right]+\sum_{m=\pm}
\sum_{k=1}^{s-1}\prod_{i=1}^{K^{\rm I }}\lambda_m(u,v_i,k)\, .
\nonumber \end{eqnarray} This transfer matrix is associated with
the canonically normalized S-matrix which is equal to unity on the
$\alg{su}(2)$ vacuum. We recall that the fundamental
representation can be realized on the space of two bosonic
variables $w^1$ and $w^2$, and two fermionic variables $\theta^3$
and $\theta^4$ \cite{Arutyunov:2008zt}. The $\su(2)$ vacuum state
is composed of a chain of $w^1$'s, {\it i.e.} $(w^1)^{\otimes
K^{\rm I}}$, where $K^{\rm I}$ is the number of excited particles
with rapidities $v_i$ and kinematic variables
$x^{\pm}_i=x\big(v_i\pm \frac{i}{g} \big)$. In the formula
(\ref{TS}) the quantities $x^{\pm}$ are the kinematic variables
corresponding to the auxiliary bound-state particle with rapidity
$u$: $x^{\pm}=x\big(u\pm \frac{i}{g}s\big)$. Finally, the
quantities $\lambda_{\pm}$ are given by
\begin{eqnarray}\nonumber \hspace{-1cm}
\lambda_\pm(u,v_i,k)&=&\frac{1}{2}\left[1-\frac{(x^-_ix^+-1)
  (x^+-x^+_i)}{(x^-_i-x^+)
  (x^+x^+_i-1)}+\frac{2ik}{g}\frac{x^+
  (x^-_i+x^+_i)}{(x^-_i-x^+)
  (x^+x^+_i-1)}\right.\\ \label{eqn;lambda-pm}
&&~~~~~~~~~~~~\left.\pm\frac{i x^+
  (x^-_i-x^+_i)}{(x^-_i-x^+)
 (x^+x^+_i-1)}\sqrt{4-\left(u-\frac{i(2k-a)}{g}\right)^2}\right]\,
 .
\end{eqnarray}
By construction, we can identify $T_{1,s}\equiv
T_{\alg{su}(2)}(u\,|\,\vec{v})$.

On the other hand, the eigenvalue of the transfer matrix on the
$\alg{sl}(2)$ vacuum, {\it i.e.} on a fermionic state
$(\theta^3)^{\otimes K^{\rm I}}$, takes the form \cite{Arutyunov:2009iq}
\begin{eqnarray}\nonumber\hspace{-0.5cm}
T_{\alg{sl}(2)}(v\,|\,\vec{u})&=&d(a,u,K^{\rm I
})\left[(1+a)\prod_{i=1}^{K^{\rm I}}\frac{x^--x^-_i}{x^+-x^-_i}
+(a-1)\prod_{i=1}^{K^{\rm
I}}\frac{x^--x^+_i}{x^+-x^-_i}\frac{x^-_i-\frac{1}{x^+}}
{x^+_i-\frac{1}{x^+}}\right.\\
&&~~~~~~~~~-\left. a\prod_{i=1}^{K^{\rm
I}}\frac{x^--x^+_i}{x^+-x^-_i}\sqrt{\frac{x^-_i}{x^+_i}}
-a\prod_{i=1}^{K^{\rm
I}}\frac{x^--x^-_i}{x^+-x^-_i}\frac{x^-_i-\frac{1}{x^+}}{x^+_i-\frac{1}{x^+}}\sqrt{\frac{x^+_i}{x^-_i}}\right]\,
, \label{TA}\end{eqnarray} where we include the following
normalization factor
\begin{eqnarray}
d(a,u,K^{\rm I })=(-1)^a\prod_{i=1}^{K^{\rm I
}}\frac{x^+-x^-_i}{x^--x^+_i}\Big(\frac{x^+_i}{x^-_i}\Big)^{\frac{a}{2}}
\prod_{n=1}^{a-1}\frac{x\big(u+\frac{2n-a}{g}i\big)-x^-_i}{x\big(u-\frac{2n-a}{g}i\big)-x^+_i}\,
.
\end{eqnarray}
The matrix $T_{a,1}$ in an anti-symmetric irrep is obtained from
$T_{\alg{sl}(2)}(v\,|\,\vec{u})$ through the replacement
$T_{a,1}\equiv T_{\alg{sl}(2)}|_{x^{\pm}\to x^{\mp},x_i^{\pm}\to
x_i^{\mp}}$.

Now we discuss the factorization of the transfer matrix which has
an auxiliary space corresponding to the long 16-dimensional irrep
with the central charge $q=1$. Recall that the transfer matrix is
defined as
\begin{align}
T_L(u|\vec{v}) = {\rm{str}}_{0}\prod^{\leftarrow}_{i>0}
\S_{0i}(u,v_i).
\end{align}
We take the auxiliary space $0$ to be the one corresponding to the long representation.
This transfer matrix acts on the tensor product
\begin{align}
V(v_1)\otimes\ldots\otimes V(v_K).
\end{align}
 It is
convenient to identify the long representation $V_{0}(u,q)$ as the
tensor product of two short $V_a(u_a(u,q))\otimes V_b(u_b(u,q))$.
Under this identification we have $\S_{0i} = \S_{ai}\S_{bi}$ and
this allows for the determination of the transfer matrix
\begin{align}
\label{superfactor} T_L(u|\vec{v}) = {\rm{str}}_{V_a\otimes V_b
}\Big(\prod^{\leftarrow}_{i>0}
\S_{ai}\Big)\Big(\prod^{\leftarrow}_{i>0}
\S_{bi}\Big)={\rm{str}}_{V_a}\Big(\prod^{\leftarrow}_{i>0}
\S_{ai}\Big){\rm{str}}_{V_b}\Big(\prod^{\leftarrow}_{i>0}
\S_{bi}\Big)
\end{align}
in terms of short representations. In the last formula the
factorization property of the supertrace operation has been used.
Obviously, the right hand side of the (\ref{superfactor})
coincides with the product
$T_{1,1}(u_a|\vec{v})T_{1,1}(u_b|\vec{v})$. This factorization
happens for any $q$. For $q=1$ it takes the form
(\ref{FactorLong}). Having proved eq.(\ref{FactorLong}), one can
substitute in (\ref{simplestHiro}) the expressions for $T_{1,2}$
and $T_{2,1}$ discussed above, and verify that the left and the
right hand sides agree with each other.

\smallskip

Higher Hirota equations have an analogous origin. For instance,
one has
\begin{eqnarray}
T^{64|2}_{L}&=&T_{1,2}^{8|1}(u+\frac{i}{g})T_{1,2}^{8|1}(u-\frac{i}{g})=T_{2,2}^{16|2}(u)
+T_{1,1}^{4|\frac{1}{2}}(u)T_{1,3}^{12|\frac{3}{2}}\, ,\\
T^{256|4}_{L}&=&T_{2,2}^{16|2}(u+\frac{i}{g})T_{2,2}^{16|2}(u-\frac{i}{g})=T_{1,2}^{8|1}(u)T_{3,2}^{16|3}(u)
+T_{2,1}^{8|1}(u)T_{2,3}^{16|3}(u)\, .
\end{eqnarray}
On the left hand side we indicate the long representations which
for generic $q$ are irreducible and can be written as the tensor
product of lower dimensional irreps. They do not have a
description in terms of the Young tableaux and for special values
of $q$ become reducible but indecomposable. All the
representations appearing on the right hand side of the Hirota
equations, like, for instance $T_{2,2}^{16|2}$, have an
associated Young tableaux.

Concrete transfer matrices in long representations  can be
obtained by using the S-matrix we have constructed in this paper.
When trying to check the above with this concrete realization,
however, one has to take into account an extra degree of freedom
corresponding to the normalization of the T-functions. This
normalization comes on top of the one chosen for the S-matrix,
which we fix in the paper to be the canonical normalization.
Although we have not studied the normalization issue in detail, we
have checked in several cases that solving for transfer matrices
of long representations from some of the Hirota equations and
plugging them in the remaining equations leads to the consistency
conditions on the transfer matrices of short representations which
are indeed satisfied.

\bibliographystyle{JHEP}
\bibliography{LitRmat}

\end{document}